%% file: ms.tex
\pdfoutput=1

\documentclass[runningheads]{llncs}
\usepackage{graphicx}
\usepackage{xspace} 
\usepackage{color}
\usepackage{url}
\usepackage{graphicx}
\usepackage{subcaption}
\usepackage{comment}
\usepackage{bbm}
\usepackage{algorithm,algorithmic}
\usepackage{amsmath}
\usepackage{amsfonts}
\usepackage{amssymb}
\usepackage{slashbox}
\usepackage{array}
\def\bfx{{{\mathbf x}}}

\def\bfz{{{\mathbf z}}}

\def\R{\ensuremath{{\mathbb R}}\xspace}

\def\X{\ensuremath{{\mathcal X}}\xspace}
\def\Y{\ensuremath{{\mathcal Y}}\xspace}

\DeclareMathOperator*{\argmin}{\arg\!\min}

\def\learner{\mathcal{A}}
\def\trset{D}
\def\cand{{\mathcal{C}}}
\def\candSize{n}
\def\candset{\mathcal{D}}
\def\trainSize{m}
\def\hypospace{\mathcal{H}}

\def\suspicion{\Psi}

\def\tx{\tilde{\bfx}}
\def\ty{\tilde{y}}
\def\cdist{Q}

\def\fspace{\mathcal{F}}
\def\mmd{\textbf{MMD}}
\def\tloss{\ell}

\def\closs{\ell}
\def\val{{D_S}}

\def\optSeq{D}
\def\tb{B}

\begin{document}
\title{Training Set Camouflage}
%
%
\author{Ayon Sen\inst{1}\and
Scott Alfeld\inst{2} \and
Xuezhou Zhang\inst{1} \and
Ara Vartanian\inst{1} \and
Yuzhe Ma\inst{1} \and
Xiaojin Zhu\inst{1}}
\authorrunning{A. Sen et al.}
%
\institute{University of Wisconsin-Madison \\
	\email{\{ayonsn, zhangxz1123, aravart, yzm234, jerryzhu\}@cs.wisc.com} \and
	Amherst College\\
	\email{salfeld@amherst.edu}}
%
\maketitle              
\begin{abstract}
We introduce a form of steganography in the domain of machine learning which we call training set camouflage.
Imagine Alice has a training set on an illicit machine learning classification task.
Alice wants Bob (a machine learning system) to learn the task.
However, sending either the training set or the trained model to Bob can raise suspicion if the communication is monitored.
Training set camouflage allows Alice to compute a second training set on a completely different -- and seemingly benign -- classification task.
By construction, sending the second training set will not raise suspicion.
When Bob applies his standard (public) learning algorithm to the second training set, he approximately recovers the classifier on the original task.
Training set camouflage is a novel form of steganography in machine learning. 
We formulate training set camouflage as a combinatorial bilevel optimization problem and propose solvers based on nonlinear programming and local search.
Experiments on real classification tasks demonstrate the feasibility of such camouflage.


\end{abstract}

\section{Introduction}
\input{intro}

\section{Training Set Camouflage}
\input{framework}


\section{Solving the Camouflage Problem}~\label{sec:opti}
\input{solvingopti}

\section{Experiments}
\label{sec:exp}
\input{clean_run_exp}

\section{Related Work}
\input{relwork}

\section{Conclusion and Discussions}
\input{discussions}

\textbf{Acknowledgment}
This work is supported in part by NSF
1545481,
1704117,
1623605,
1561512,
and the MADLab AF Center of Excellence FA9550-18-1-0166.

%
%
%
 \bibliographystyle{splncs04}

 %

 \vfill 
\pagebreak
 \section{Appendix A: MMD as Eve's Detection Function}
 \input{mmd}

 
\end{document}

%% file: intro.tex

\begin{figure*}[h]
	\centering
	\begin{subfigure}[t]{0.9\textwidth}
		\centering
		\includegraphics[width=\textwidth]{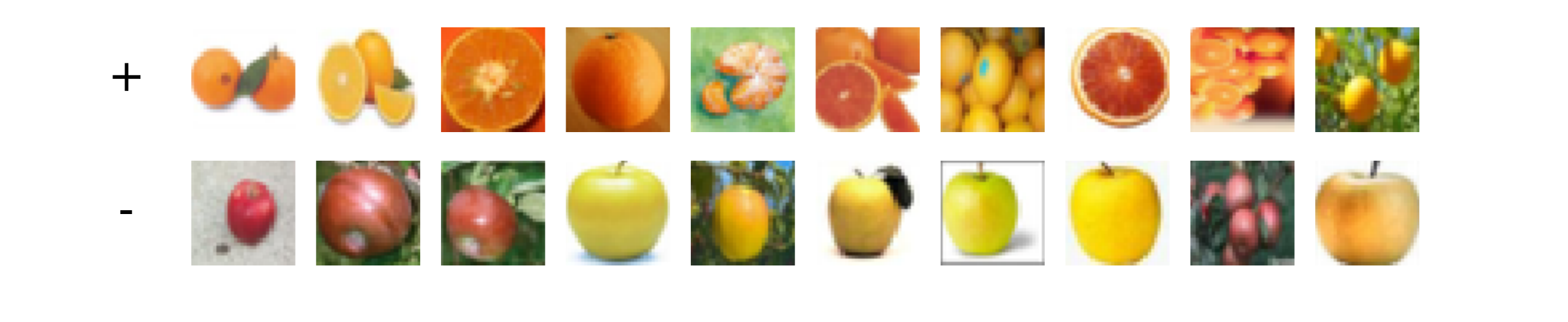}
		\caption{Camouflaged training set}~\label{fig:cts}
	\end{subfigure}
	
	\begin{subfigure}[t]{0.9\textwidth}
		\centering
		\includegraphics[width=\textwidth]{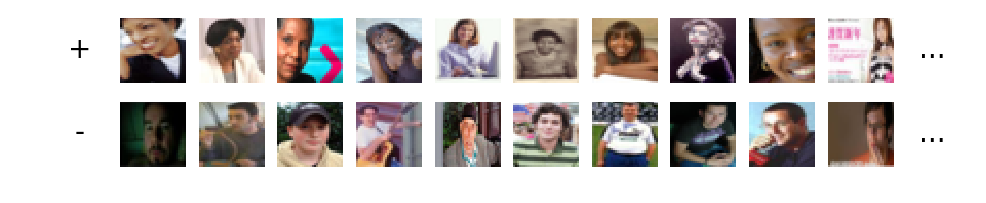}
		\caption{Secret classification task}~\label{fig:sd}
	\end{subfigure}
	\caption{Example of training set camouflage}~\label{fig:example}
\end{figure*}
Look at the classification training set shown in Figure~\ref{fig:cts}.
The top row contains instances of class positive (+), and the bottom shows instances of class negative (-).
These images can be fed into a machine learner to learn a model which will successfully classify future, previously unseen instances (images) as + or -.
If you think that the task is fruit image classification (orange vs. apples) then you have already been successfully fooled, in a sense to be made precise below.
The actual intended task is to classify woman vs. man, with samples shown in Figure~\ref{fig:sd}.
Indeed, a standard logistic regression learner~\cite{hosmer2013applied} trained on only the images in Figure~\ref{fig:cts} achieves high gender classification accuracy on the images in Figure~\ref{fig:sd}.

In this paper, we consider an agent Alice who has a secret classification task (e.g., classifying images of women and men) and a corresponding private training set (women and men images).
Alice wants to train a second agent, Bob, on the secret task.
However, the communication channel between them has an eavesdropper we refer to as a third agent Eve.
Eve takes the role of a data verifier, who will terminate communication (and refuse to deliver the data to Bob) if she is suspicious of what Alice is sending.
Sending the private training set would reveal Alice's intention; sending the model parameters directly will also raise suspicion.
Alice must camouflage the communication for it to look mundane to Eve, while avoiding excessive coding tricks with Bob beforehand.
In the present work, we show how Alice can construct a camouflaged training set on a {\em cover task}  which (i) does not look suspicious to Eve, and (ii) results in Bob learning an accurate model for the secret task. 
In the previous example, Eve noticed that Alice sent images of apples and oranges which seems benign, and knew nothing of Alice's secret task of women vs men.




Hiding information in plain sight such that its presence is not suspected is known as steganography.
Steganography is not new. 
In the fifth century BCE  messengers would have their heads shaved and a message written on their scalp.
Regrowing their hair served to hide the message which would only be revealed because the intended recipient knew to shave the messenger's head~\cite{krenn2004steganography}.
In more modern times, steganographic techniques are used to detect unauthorized distribution of digital media~\cite{cox2005information}.

Note that, steganography is different from crypotgraphy~\cite{katz1996handbook,van2014encyclopedia}, where the goal is to hide the data content. 
In cryptography, the communicating agents have access to some particular key (pairs) which is used to encrypt and decrypt data. 
Cryptography cannot be used if someone monitoring the data can alter the data or stop the data transmission entirely. 
In such cases, steganography becomes important because we do not want any intervening eavesdropper to become suspicious and stop the data transmission.

The role and capabilities of the eavesdropper are key in selecting how to hide information.
Eve can be either passive and merely observes traffic~\cite{cachin1998information}, or active and tries to modify the hidden message~\cite{queirolo2011steganography,chandramouli2002mathematical}.
In this manuscript we assume a passive observer, whose only ability is to refuse to deliver Alice's message. 
To our knowledge, steganography for machine learning in this context is new.
In the area of adversarial learning~\cite{biggio2017wild}, however, much work has been done investigating how an agent can assert control over a learner by manipulating input data.
We note that training set camouflage differs from so called training-time or ``poisoning attacks"~\cite{kloft2010online} in two primary ways: (i) Alice aims to communicate information to Bob about a potentially completely unrelated task, not affect his behavior on the original task and (ii) Alice specifically aims to avoid detection by Eve.



Due to the widespread use of machine learning in sensitive fields ranging from social media to health care, the study of the security ramifications of using ML techniques is well studied \cite{barreno2010security,lowd2005adversarial}.
The work presented herein adds to this conversation, as we reveal an additional avenue of attack. 
For example, Bob might be a model that classifies job applicants as ``should hire" and ``shouldn't hire".
The company may have many records (collected over years) of job applicants and how they performed.
It is expected from Alice to select a subset of these records and present to Bob, with the idea that training on the complete set is too time consuming.
But Alice may be a malicious agent and wants Bob to actually learn some additional bias (e.g., racial, gender etc.).
In such a scenario, Alice will select a subset of records that satisfies her goals while Eve's responsibility is to verify the data sent by Alice to Bob.
Our specific contributions in this paper are as follows: (i) We propose a general mathematical framework for defining how Alice can achieve training set camouflage.
(ii) We formulate a nonlinear-program based approach for performing Alice's task for a general class of learner (Bob) and eavesdropper (Eve), and two combinatorial-search based approaches for arbitrary learners/eavesdroppers.


%% file: framework.tex
In this section we describe the three agents Bob, Alice and Eve,  
and formulate a camouflage optimization problem for Alice, parametrized by Bob and Eve's definitions.


The agent Bob uses a standard learning algorithm $\learner: \candset\mapsto\hypospace$ which, given a training set $D$, learns a hypothesis \(\learner(D)\) in a hypothesis space \(\hypospace\).
The resulting hypothesis maps instances in the input space \X to the output space \Y.
This can be multi-class classification (three or more classes) or regression, though in the present work we focus on binary classification.
We assume that Bob's learning algorithm is ``open source''.
That is, all information about $\learner$ is known to all agents.
However, Bob and Alice have shared knowledge on class naming: which class is positive and which negative.  
For $K$-class classification this shared knowledge requires $O(K\log K)$ bits, as Alice must communicate a mapping from \(K\) classes to \(K\) classes.
For example, when Alice sends Bob orange and apple images for the secret task of woman vs man, Alice must communicate to Bob whether orange maps to woman and apple to man, or vice versa.


Alice is an agent who wants to train Bob.
She has a secret classification task and the corresponding private dataset \(\val\).
In addition, she has access to a public pool of \(\candSize\) instances \(\cand=\{(\bfx_i, y_i)_{1:\candSize}\}\) (the {\em camouflage pool}) drawn i.i.d. from \(\mathbb{\cdist}_{(\bfx, y)}\) which we call the {\em cover data distribution}.
Note that this is not the  distribution from which \(\val\) is drawn.
In the preceding example, \(\mathbb{\cdist}_{(\bfx, y)}\) is the distribution over orange and apple images, whereas \(\val\) is a collection of photographs of women and men.

Alice seeks to select a camouflaged training set $D \subset \cand$ which she will send to Bob for training. 
Alice wants Bob to succeed on the secret task, thus she seeks to find a \(D\) which minimizes the risk of Bob's resulting model: 
\begin{align*}
\mathcal{L}_\learner(D) = \frac{1}{|\val|}\sum_{(\tx, \ty)\in \val}\tloss(\learner(D)(\tx), \ty)
\end{align*}
where \(\tloss(\hat y, y)\) is a loss function.
Note the private $D_S$ is used to define the quality of $D$.
Also, we note that $D_S$ and $D$ are in different domains.



Eve is a passive eavesdropper.
She observes the dataset $D$ which Alice sends to Bob, but does not alter it.
Her goal is to determine if $D$ is normal or suspicious.
Specifically, Eve is suspicious of any data unless it appears to be an i.i.d. sample drawn from  \(\mathbb{\cdist}_{(\bfx, y)}\).
She has access to the camouflage pool \(\cand\), but like Alice she cannot draw additional instances from \(\mathbb{\cdist}_{(\bfx, y)}\).
Eve uses \(\cand\) and a detection function $\suspicion:\candset\times\candset\mapsto\mathbb{R}_+$ to examine \(D\).
Here $\candset$ is the domain of all possible training sets.
She flags \(D\) as suspicious if \( \suspicion(\cand, D)\ge0 \).
It is useful to think of $\suspicion$ as a two-sample statistical test that aims to determine whether $\cand, D$ come from the same underlying distribution.
We envision Alice as Eve's adversary.
As with Bob, we assume that Eve is ``open source''.
Namely, Alice can see Eve's detection function $\suspicion$.



The agents Alice and Eve have conflicting goals.
Alice aims to secretly communicate to Bob by way of constructing a training set, while Eve is suspicious and will flag any data sent to Bob if it looks suspect.
Bob dutifully learns a model and for our purposes has no explicit intentions.
We make explicit each Agent's knowledge in Table~\ref{tab:agentInfo}.
\newcolumntype{P}[1]{>{\centering\arraybackslash}p{#1}}
\begin{table*}[htb]
	\centering
	\renewcommand{\tabcolsep}{1pt}
	\caption{Information available to different agents}~\label{tab:agentInfo}
	\resizebox{\textwidth}{!}{%
		\begin{tabular}{|c|P{2.0cm}|P{2.0cm}|P{2.0cm}|P{2.0cm}|P{2.5cm}|}
			\hline
			
			\textbf{Agent} & \textbf{Secret Set} & \textbf{Camouflage Pool} & \textbf{Bob's Learner} & \textbf{Detection Function} & \textbf{Camouflaged Training Set}  \\
			& $\val$ & $\cand$ & $\learner$ & $\suspicion$ &  $\optSeq$\\\hline
			Bob & No & Yes/No & Yes & Yes/No & Yes \\\hline
			Alice & Yes & Yes & Yes & Yes & Yes \\\hline
			Eve & No & Yes & Yes & Yes & Yes \\\hline
		\end{tabular}
	}
\end{table*}

\begin{figure}[!htb]
	\centering
		\includegraphics[width=0.9\textwidth]{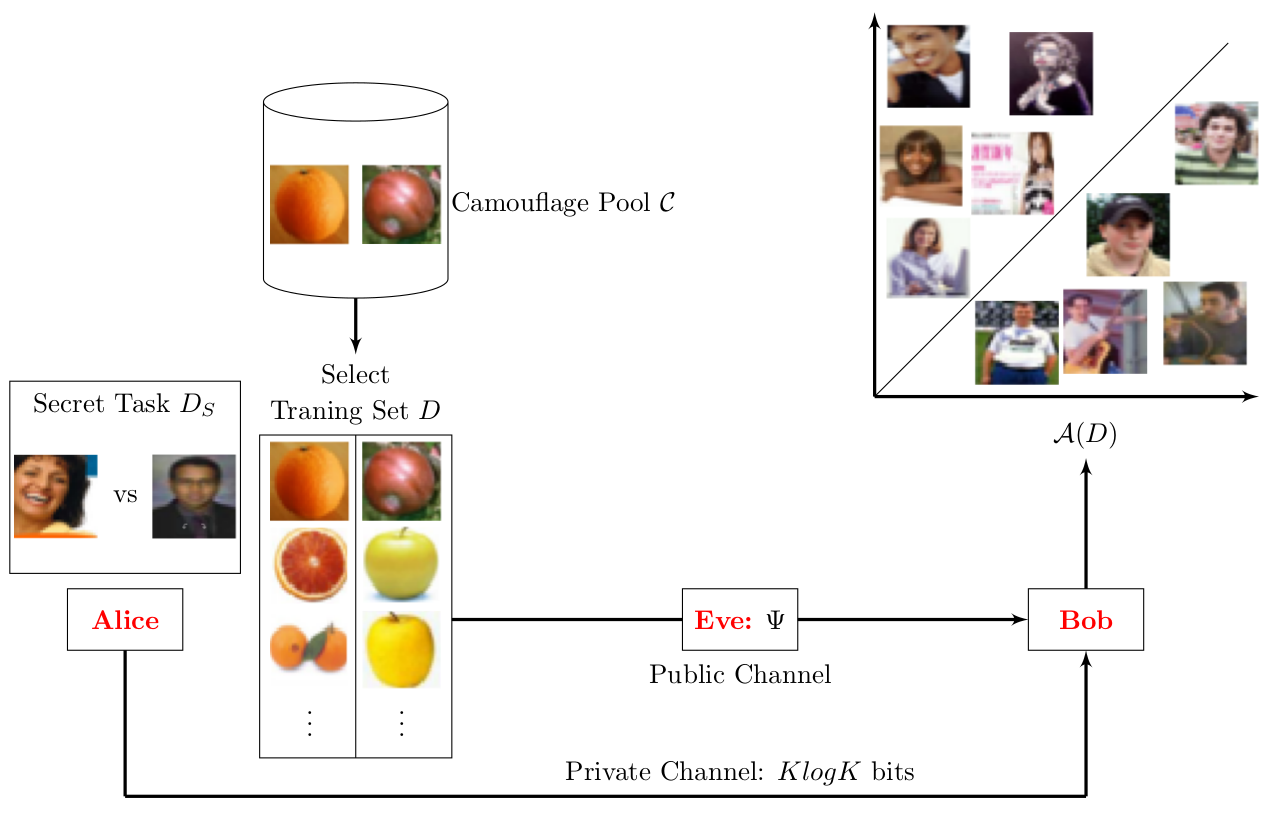}
	\caption{Training set camouflage framework. We show the three agents along with the classification task, camouflage pool, camouflage training set and Eve's detection function}~\label{fig:framework}
\end{figure}

With the agents defined, we can now formulate Alice's goal:
\begin{align}
\argmin_{D\subset \cand} &  \frac{1}{|\val|}\sum_{(\tx, \ty)\in \val}\tloss(\learner(D)(\tx), \ty)\nonumber\\ 
\text{ s.t.
} & \suspicion(\cand, D)<0 
\label{aliceopti}
\end{align}
That is, she seeks a camouflaged training set \(D\) from the cover data pool.
$D$ should not be flagged as suspicious by Eve.
$D$ should also make Bob learn well, similar to as if Alice directly gave Bob her private data set \(\val\).
An example of the training set camouflage in action is shown in Figure~\ref{fig:framework}.

%% file: solvingopti.tex
In this section, we propose three methods of solving the optimization problem defined in (\ref{aliceopti}).
We first show how the optimization problem can be reduced to a nonlinear programming problem for a broad class of learners.
We relax the resulting optimization problem to one which is computationally efficient to solve.
We then present two combinatoric methods as heuristic methods applicable to any learner.

\subsection{Nonlinear Programming (NLP)}\label{sec:NLP}

We assume Bob's machine learning algorithm $\learner$ solves a convex optimization problem.
Specifically, Bob performs  regularized empirical risk minimization.
This covers a wide range of learners such as support vector machines~\cite{hearst1998support}, logistic regression~\cite{hosmer2013applied}, and ridge regression~\cite{hoerl1970ridge}.
Let $\Theta$ be Bob's hypothesis space,  $\closs$ his loss function, and $\lambda$ his regularization parameter, respectively.
Let $\trainSize:=\vert \trset \vert$ be given.
We convert Alice's optimization problem~(\ref{aliceopti}) into a nonlinear programming problem as follows.

\textbf{Step 1.} 
Using the definition of Bob, we rewrite~(\ref{aliceopti}) as
\begin{align}
	\min_{\trset\subset \cand, \hat\theta \in \Theta}& ~~&&\frac{1}{\vert \val \vert} \sum_{(\tx, \ty)\in \val}\tloss(\hat{\theta}, \tx, \ty)\nonumber\\
	\text{s.t.}& ~~&&\hat{\theta} = \argmin_{\theta \in \Theta} \sum_{(\bfx, y)\in\trset} \closs(\theta, \bfx, y) + \frac{\lambda}{2} \Vert \theta \Vert^2\nonumber\\
	& &&\suspicion(\cand,D) < 0, \nonumber \\
	& &&\vert \trset \vert = \trainSize.\label{eq:MMDConstraint}
\end{align}
We make note that in both levels of this bilevel optimization problem (the upper and lower levels corresponding with Alice and Bob, respectively) \(\tloss(\cdot)\) is being minimized.
That is, Alice and Bob both seek to minimize the loss of Bob's resulting model.
Due to its combinatorial nature, this is a computationally difficult problem to solve.

\textbf{Step 2.}
Since Bob's learning problem (the lower level optimization problem) is assumed to be convex, satisfying its Karush-Kuhn-Tucker (KKT) conditions is necessary and sufficient for a point to be optimal~\cite{wu2007karush,liu2016teaching}.
Thus we replace the lower level optimization problem in~(\ref{eq:MMDConstraint}) with the KKT conditions to obtain a single-level optimization problem:

\begin{align}
	\min_{\trset\subset \cand, \hat\theta \in \Theta}& ~~&&\frac{1}{\vert \val \vert} \sum_{(\tx, \ty)\in \val}\tloss(\hat{\theta}, \tx, \ty)\nonumber\\
	\text{s.t.}& ~~&& \sum_{(\bfx, y)\in\trset} \triangledown\closs(\hat\theta, \bfx, y) + \lambda\hat\theta = 0,\nonumber\\
	& &&\suspicion(\cand,D) < 0, \nonumber\\
	& && \vert D \vert = \trainSize.
\end{align}
While now a single level optimization problem, selecting a subset $\trset\subset \cand$ is still a combinatorial problem and computationally expensive to solve.
In what comes next we relax this problem to one of continuous optimization.

\textbf{Step 3.}
We introduce binary indicator variable $b_i$ for each instance $(\bfx_i, y_i) \in \cand$.
A value of \(1\) indicates that the instance is a member of the training set \(D\).
Also dropping the hat on $\hat\theta$ for simplicity.
This yields:
\begin{align}
	\min_{\theta \in \Theta;b_1,\ldots,b_{|\cand|};b_i \in \{0, 1\}} ~~\frac{1}{\vert \val \vert} \sum_{(\tx, \ty)\in \val}{\tloss(\theta, \tx, \ty)} \nonumber\\
	\text{s.t.} ~~\sum_{i=1}^{\candSize} b_i\triangledown\closs(\theta, \bfx_i, y_i) + \lambda\theta = 0,\nonumber\\
	\suspicion(\cand,\{b_i(\bfx_i, y_i)|(\bfx_i,y_i)\in\cand,b_i\ne 0\}) < 0\nonumber\\
	\sum_{i=1}^{\candSize}b_i = \trainSize.
\end{align}
This is known as a Mixed Integer Non-Linear Optimization Problem (MINLP)~\cite{bussieck2003mixed}.
MINLP problems are generally hard to solve in practice. 
However, phrasing the problem in this way yields a natural relaxation.
Namely we relax $b_i$ to be continuous in $[0,1]$, resulting in the following non-linear optimization problem:
\begin{align}
	\min_{\theta \in \Theta;b_1,\ldots,b_{\candSize} \in [0, 1]} ~~\frac{1}{\vert \val \vert} \sum_{(\tx, \ty)\in \val}{\tloss(\theta, \tx, \ty)} \nonumber\\
	\text{s.t.} ~~\sum_{i=1}^{\candSize} b_i\triangledown\closs(\theta, \bfx_i, y_i) + \lambda\theta = 0,\nonumber\\
	\suspicion(\cand,b_1,\ldots,b_{|\cand|}) < 0,\nonumber\\
	\sum_{i=1}^{\candSize}b_i = \trainSize.
\end{align}
Note that in this equation we scale the gradient of the loss function for each $(\bfx_i, y_i)$ by the corresponding $b_i$.
This $b_i$ indicates the importance of an instance in the training set.
In essence, the learner is training on a ``soft'' version of the dataset, where each training example is weighted.
Similarly, when calculating the detection function we weigh each instance in the training set by its corresponding $b_i$.
The exact nature of this weighing depends on the detection function itself.
We further note that the nonlinear optimization problem is non-convex.
As such, Alice must seed her solver with some initial $\{b_i\}$.
This is discussed further in Section~\ref{sec:exp}.

After solving this (continuous) optimization problem, Alice must round the $\{b_i\}$'s into binary indicators so that she can select a training set to send to Bob.
Alice uses a rounding procedure that proposes $m+1$ candidate training sets $D^{(1)}, \ldots, D^{(m+1)}$ from the continuous solution $\{b\}$.
The candidate training sets include 
(1) the training set $D^{(1)}$ consisting of the $m$ items with the largest $b$ values, 
(2) the seed training set before running optimization, 
(3) $m-1$ other training sets that ``interpolate'' between 1 and 2.
Alice then checks $D^{(1)}, \ldots, D^{(m+1)}$ for feasibility (satisfying $\suspicion$) and picks the best one. 
Note the seed training set is feasible, hence Alice is guaranteed to have a solution.
The interpolation scheme ensures that Alice will find a solution no worse than the seed set.

Concretely, let $S$ be the $m$-item seed training set and $\cand \backslash S$ be the remaining items.
Alice sorts items in $S$ by their $b$ values.
Separately, Alice sorts items in $\cand \backslash S$ by their $b$ values.
Then, Alice starts from $S$ and sequentially swaps the least-valued item in $S$ with the largest-valued item in $\cand \backslash S$.
She performs $m$ swaps.  This produces the $m+1$ candidate training sets, including the original $S$.
It can be shown that the $m$ items with the largest $b$ values will be one of the training sets.


\subsection{Uniform Sampling}

For any learner Bob, even one which does not solve a convex empirical risk minimizing problem discussed above, Alice has a simple option for finding a training set.
Let Alice's budget \(\tb\) denote the number of times Alice is able to train the classifier $\learner$.
She first creates \(\tb\) training sets $D^{(1)}, \ldots, D^{(\tb)}$, each by sampling $m$ points uniformly without replacement from her camouflage pool $\cand$, such that each $D^{(j)}$ successfully bypasses Eve i.e., $\suspicion(\cand, D^{(j)})<0$.
Among these $B$ training sets, she then picks the $D^{(j)}$ with the lowest objective value in~\eqref{aliceopti}.
This procedure captures what Bob would learn if given each feasible training set.

\subsection{Beam Search}~\label{sec:search}

We now describe a heuristic beam search algorithm~\cite{rich1991artificial} to approximately solve Alice's optimization problem~\eqref{aliceopti}.
This process is similar to uniform sampling, described above, but instead of independently generating a new training set every time, Alice performs a local search to augment a proposed training set incrementally.

The state space consists of all training sets $\optSeq\subset \cand$ such that $\vert \optSeq \vert = \trainSize$ and $\suspicion(\cand, \optSeq) < 0$. 
Two training sets that differ by one instance are considered neighbors. 
For computational efficiency, we do not consider the entire set of neighbors at each step.
Instead, we evaluate a randomly selected subset of neighbors for each training set in the beam.
The beam \(\mathcal{D}\) is initialized by selecting \(w\) training sets at random.
The width ($w$) of the beam is fixed beforehand.
From the union of evaluated neighbors and training sets in the current beam, we select the top $w$ training sets (based on the value of the objective function in (\ref{aliceopti})) to reinitialize the beam and discard the rest.
Note that training sets which would be flagged by Eve are not present in the statespace (because Alice has full knowledge of Eve, she need not consider any set that Eve would reject).
We continue the search process until a pre-specified search budget $B$ (number of times the classifier $\learner$ is trained) is met.
Algorithm~\ref{alg:BeamSearch} shows the search procedure with random restarts.


\begin{algorithm}[htb]
	\caption{Beam Search for Solving the Camouflage Problem}\label{alg:BeamSearch}
	\begin{algorithmic}[1]
		\STATE{Input: Camouflage Pool: $\cand$, Risk: $\mathcal{L}_\learner$, Beam Width: $w$, Budget: $B$, Neighborhood Function: $\mathcal{N}$, Size: $\trainSize$, Detection Function: $\suspicion$, Restarts: $R$ }
		\FOR{$r = 1 \to R$ }
		\STATE{$\mathcal{D}\leftarrow$ $w$ randomly selected subsets of size $\trainSize$ from $\cand$ such that $\suspicion(\cand,D)<0$}
		\WHILE{budget $B/R$ not exhausted}
			\STATE{$\mathcal{D} \leftarrow \mathcal{D} \cup \mathcal{N}(\mathcal{D},\cand, \suspicion)$, the neighbors}
			\STATE{$\mathcal{D}\leftarrow w$ training sets from $\mathcal{D}$ with smallest $\mathcal{L}_\learner(D)$ values}
		\ENDWHILE
		\ENDFOR
		\RETURN{the best $D$ found within total budget}
	\end{algorithmic}
\end{algorithm}

%% file: clean_run_exp.tex
We investigated the effectiveness of training set camouflage through empirical experiments on real world datasets. 
Our results show that camouflage works on a variety of image and text classification tasks: 
Bob can perform well on the secret task after training on the camouflaged training set, and the camouflaged training set passes Eve's test undetected.
We start by discussing the three agents.

\textbf{Bob}.  We considered the logistic regression learning algorithm for Bob.
Logistic regression is a popular learner and is regularly used in practice. 
Bob set the weight of the regularization parameter to $1$. 

\textbf{Eve}.
The training set camouflage framework is general with respect to Eve's detection function.
For our experiments we used Maximum Mean Discrepancy  ({\mmd})~\cite{gretton2012kernel} as the core of Eve's detection function.
We used {\mmd} as it is a popular and widely used two-sample test~\cite{dziugaite2015training}.
Unfortunately {\mmd} cannot be directly applied to the camouflage framework as its application requires that the two samples have the same size.
We introduce {\mmd} and how Eve used it in Appendix A.
The level-$\alpha$ for this detection function was set to $0.05$ (i.e., the probability of incorrectly rejecting a benign training set is 5\%). 

\textbf{Alice}. 
We considered three different Alices. Each of them used one of the proposed solvers.
For each secret task, Alice had access to multiple camouflage candidate tasks.
Alice can run her solver on each of these tasks separately and then select the best one, but this would be time consuming and thus instead she started by identifying a suitable camouflage task.
For this purpose, all three Alices used uniform sampling (as this is the easiest algorithm to implement, and makes the weakest assumptions) with a search budget of $80,000$ (divided equally among candidate tasks).
This meant that Alice stopped after training the logistic regression learner $80,000$ times.
For each candidate task Alice identified a training set using this budget.
Then she selected the best task (as her cover task) based on the loss on the secret set.

Next, all three Alices used their respective solvers (NLP, beam search and uniform sampling) to find a camouflaged training set.
We assumed that all of them were allotted a fixed amount of time for this purpose.
This time was set as the time required to run the NLP solver. 

The Alice who used the NLP solver seeded the solver with the camouflaged training set found during the candidate task identification phase.
The Alice who used the beam search solver performed random restarts each with a per-restart budget of $B/R = 16,000$.
Here the width of the beam was $w=10$ and for each training set in the beam, $50$ randomly selected neighbors were evaluated during each iteration.
It should be noted that both beam search and uniform sampling are stochastic in nature.
We run the Alices who used these solvers five times. 
We then report the average.
Alice constructed camouflaged training sets of size $m=2, 20$ and $50$, $100$ and $500$, and set the loss $\ell$ to logistic loss with natural logarithm.
All experiments were run on an Intel(R) Core(TM) i7-7700T CPU @2.90GHz machine, using one thread.

\textbf{Evaluation metrics}.
As is standard to estimate generalization performance of a learned model, we used a separate test set, generated from the same distribution as the secret set $D_S$ and not known to any agent, to estimate Bob's generalization error when trained on Alice's camouflaged training set $D$.
We compare these values to two additional quantities:
(``random'') when Bob is trained on a uniform sample of size $m$ from the cover data distribution, which we expect to perform poorly; and 
(``oracle'') when Bob is trained directly on Alice's secret set $D_S$, ignoring Eve's presence.
The oracle gives us an estimate on how much performance Bob is losing due to using the camouflage framework to fool Eve.

\subsection{Datasets}
We performed experiments for four secret tasks: WM (CIFAR-100~\cite{krizhevsky2009learning}), GP (OpenImages~\cite{openimages}), CA (20-newsgroups~\cite{joachims1996probabilistic}) and DR (All The News dataset~\cite{allthenews2017}).
The two letters in the acronym represent the two classes in the corresponding task (see Table~\ref{tab:dataSum}).
The first two tasks were image classification while the remaining two were text classification.
For the image tasks we selected eight candidate cover tasks. 
Six of them were from the MNIST handwritten digits: 17, 71, 25, 52, 69 and 96.
The other two were from the CIFAR-100 dataset: OA and AO.
Similarly for the text tasks we also selected eight candidate cover tasks.
All of them were from the 20-newsgroups dataset: BH, HB, IM, MI, AM, MA, MX and XM.
As before the acronyms here represent the class names.

\begin{table}
	\centering
	\caption{Summary of secret sets and camouflage pools.}~\label{tab:dataSum}
		\begin{tabular}{|c|c|c|c|c|c|c|}
		\hline
		\textbf{Dataset} & \textbf{Type} & \textbf{\# Features} & \textbf{class 1} & \textbf{class 2} & \textbf{\# class 1} & \textbf{\# class 2} \\\hline
		WM & Image & 2048 & \textbf{w}oman & \textbf{m}an & 500 & 500 \\\hline
		GP & Image & 2048 & hand\textbf{g}un & \textbf{p}hone & 400 & 400 \\\hline
		CA & Text & 300 & \textbf{c}hristian & \textbf{a}theist & 599 & 480 \\\hline
		DR & Text & 300 & \textbf{d}emocratic & \textbf{r}epublican & 800 & 800 \\\hline
		17 & Image & 2048 & digit \textbf{1} & digit \textbf{7} & 600 & 600 \\\hline
		25 & Image & 2048 & digit \textbf{2} & digit \textbf{5} & 600 & 600 \\\hline
		69 & Image & 2048 & digit \textbf{6} & digit \textbf{8} & 600 & 600 \\\hline
		OA & Image & 2048 & \textbf{o}range & \textbf{a}pple & 600 & 600 \\\hline
		BH & Text & 300 & \textbf{b}aseball & \textbf{h}ockey & 994 & 999 \\\hline
		IM & Text & 300 & \textbf{i}bm & \textbf{m}ac & 982 & 963 \\\hline
		AM & Text & 300 & \textbf{a}utos & \textbf{m}otorcycles & 990 & 996 \\\hline
		MX & Text & 300 & \textbf{m}s-windows & windows \textbf{x} &985 & 988 \\\hline
	\end{tabular}
\end{table}

For images we used ResNet~\cite{he2016deep} to generate feature vectors of dimension $2048$. 
For this purpose we removed the output layer and used the values found in the penultimate layer of the network.
For text we used Word2Vec~\cite{mikolov2013efficient} to generate feature vectors of dimension $300$ by averaging over the word vectors in an article.
We also removed punctuation and stop words before generating the word vectors.
A summary of the secret sets and camouflage pools can be found in Table~\ref{tab:dataSum}. As mentioned previously, we kept a held out test set for each of the secret tasks. The number of class 1 and class 2 instances were 100/100, 100/100, 398/319 and 200/200 respectively for WM, GP, CA and DR.
Here the two numbers (num1/num2) represent the number of instances in class1 and class2 respectively.

\begin{table}[!htb]
	\centering
	\caption{Logistic loss ($\frac{1}{|D_S|}\sum_{(\tx,\ty)\in D_S}\log(1+\exp(-\ty w^\top\tx))$) after performing Uniform Sampling search with search budget $10,000$ for image secret tasks. The best results for each secret task is shown in bold.}~\label{tab:imageCandSelect}
		\begin{tabular}{|c|c|c|c|c|c|c|c|c|c|}
		\hline
		m & \backslashbox{Secret}{Camouflage}
		&\makebox[2em]{17}&\makebox[2em]{71}
		&\makebox[2em]{25}&\makebox[2em]{52} &\makebox[2em]{69}&\makebox[2em]{96}
		&\makebox[2em]{OA}&\makebox[2em]{AO}\\\hline 
		2 & WM & $0.671$ & $0.631$ & $0.643$ & $0.638$ & $0.671$ & $0.640$ & $\mathbf{0.606}$ & $0.647$    \\\cline{2-10}
		& GP & $0.481$ & $0.541$ & $0.458$ & $\mathbf{0.443}$ & $0.516$ & $0.463$ & $0.541$ & $0.558$ \\\hline
		20 &WM & $0.790$ & $0.611$ & $0.672$ & $0.688$ & $0.798$ & $0.679$ & $\mathbf{0.584}$ & $0.731$ \\\cline{2-10}
		& GP & $0.480$ & $0.510$ & $0.433$ & $0.390$ & $0.632$ & $\mathbf{0.337}$ & $0.510$ & $0.531$ \\\hline
		50 & WM & $0.874$ & $0.614$ & $0.705$ & $0.772$ & $1.116$ & $0.802$ & $\mathbf{0.606}$ & $0.856$ \\\cline{2-10}
		& GP & $0.565$ & $0.479$ & $0.473$ & $\mathbf{0.387}$ & $1.047$ & $0.421$ & $0.479$ & $0.506$   \\\hline
		100 & WM & $0.939$ & $0.651$ & $0.796$ & $0.868$ & $1.441$ & $1.080$ & $\mathbf{0.620}$ & $0.907$ \\\cline{2-10}
		& GP & $0.589$ & $0.469$ & $0.639$ & $\mathbf{0.439}$ & $1.157$ & $0.530$ & $0.0.970$ & $0.538$  \\\hline
		500 & WM & $1.193$ & $0.794$ & $1.107$ & $1.190$ & $2.688$ & $1.988$ & $\mathbf{0.705}$ & $1.283$    \\\cline{2-10}
		& GP & $0.955$ & $\mathbf{0.577}$ & $1.449$ & $0.714$ & $1.795$ & $0.862$& $1.416$ & $0.800$   \\\hline
	\end{tabular}
\end{table}

\begin{figure}[!htb]
	\centering
	\begin{subfigure}[t]{0.85\textwidth}
		\centering
		\includegraphics[width=\textwidth]{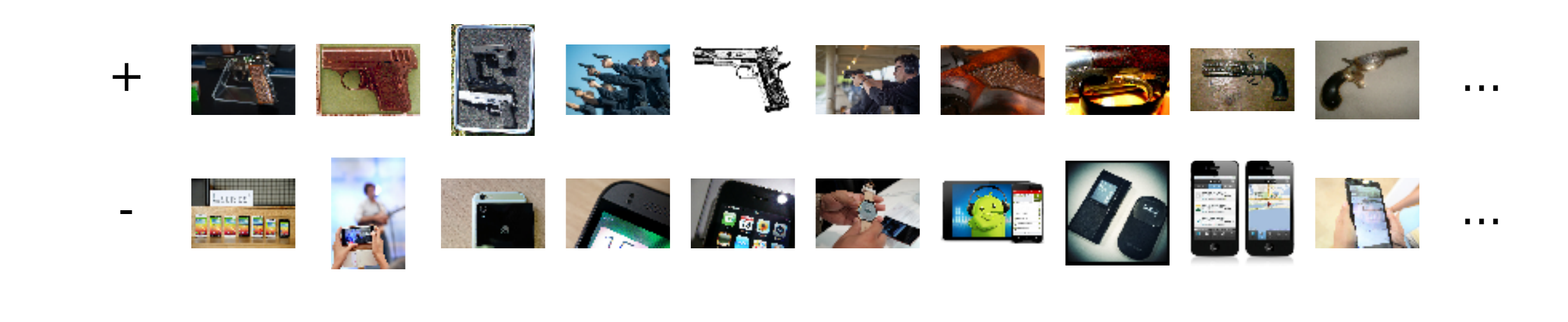}
		\caption{Samples from the secret task Handgun vs. Phone (GP)}~\label{fig:GP}
	\end{subfigure}
	
	\begin{subfigure}[t]{0.85\textwidth}
		\centering
		\includegraphics[width=\textwidth]{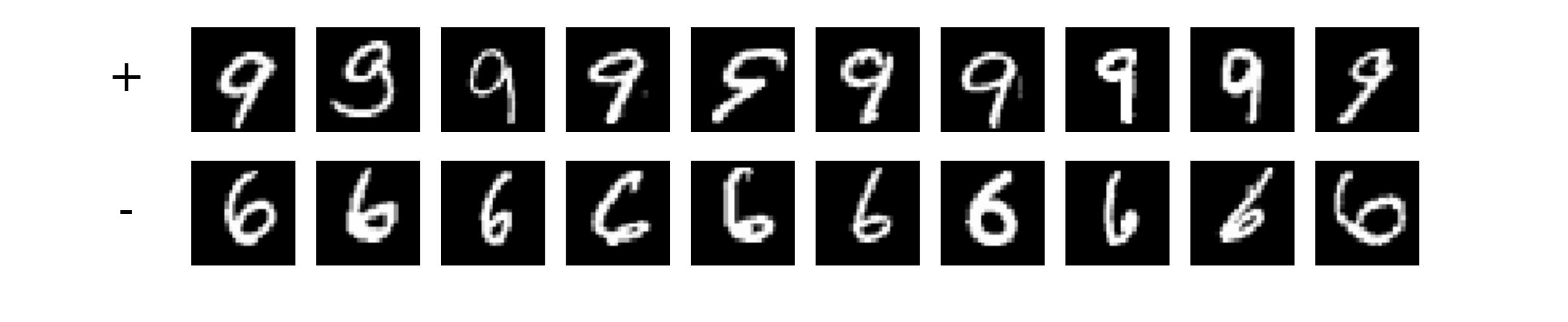}
		\caption{Camouflaged training set using 9 vs. 6}~\label{fig:GP96}
	\end{subfigure}
	
	\begin{subfigure}[t]{0.85\textwidth}
		\centering
		\includegraphics[width=\textwidth]{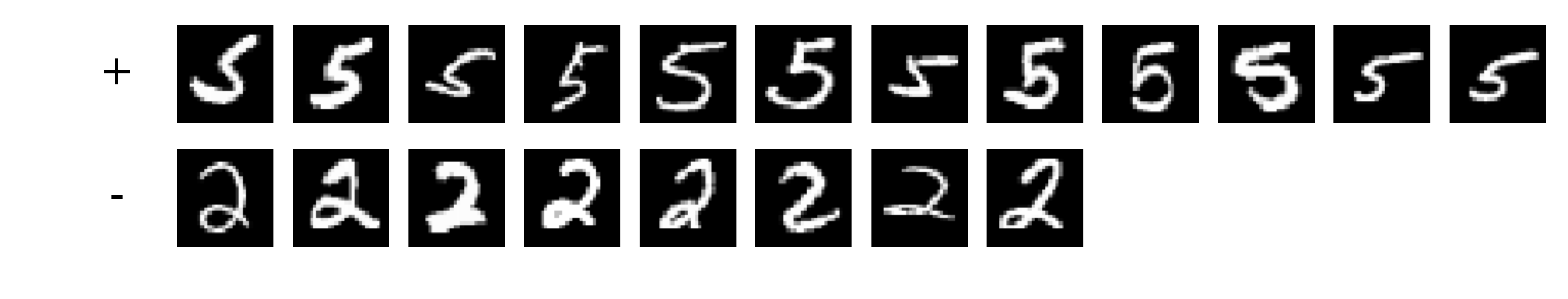}
		\caption{Camouflaged training set using 5 vs 2}~\label{fig:GP52}
	\end{subfigure}
	\begin{subfigure}[t]{0.85\textwidth}
		\centering
		\includegraphics[width=\textwidth]{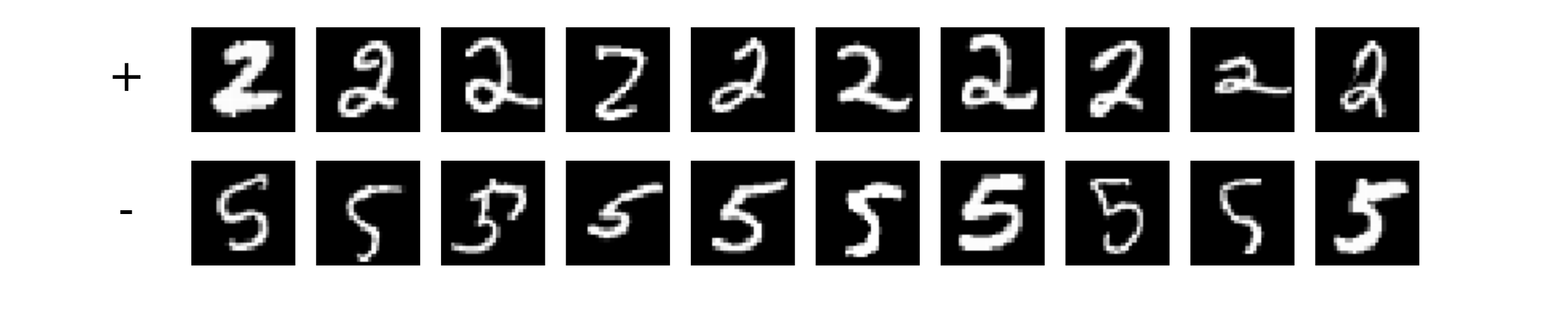}
		\caption{Camouflaged training set using 2 vs. 5}~\label{fig:GP25}
	\end{subfigure}
	\caption{Samples of GP secret set, and the top three camuflaged training set found during the candidate selection phase for $m=20$.}~\label{fig:candidateCover}
\end{figure}

Alice first selected a suitable camouflage task for each of the secret tasks.
For each candidate task she used a search budget of $10,000$ (for a total of $80,000$ budget).
The results of this phase are shown in Table~\ref{tab:imageCandSelect} and \ref{tab:textCandSelect}.
The selected camouflage tasks are shown in Table~\ref{tab:tasks}.
It should be noted that the logistic error reported in the tables are large ($>0.693$) in some cases indicating that some of these cover tasks will perform worse than random chance on secret tasks.
However, this was not true for the selected cover tasks.
The top three camouflaged training sets for GP ($m=20$) identified during this phase are shown in Figure~\ref{fig:candidateCover}.

\begin{table}[!htb]
	\centering
	\caption{Logistic loss after performing Uniform Sampling search with search budget $10,000$ for text secret tasks. The best results for each secret task is shown in bold.}~\label{tab:textCandSelect}
	\begin{tabular}{|c|c|c|c|c|c|c|c|c|c|}
		\hline
		m & \backslashbox{Secret}{Camouflage}
		&\makebox[2em]{BH}&\makebox[2em]{HB}
		&\makebox[2em]{IM}&\makebox[2em]{MI} &\makebox[2em]{AM}&\makebox[2em]{MA}
		&\makebox[2em]{MX}&\makebox[2em]{XM}\\\hline 
		2 & CA & $0.6845$ & $0.6846$ & $0.6868$ & $0.6862$ & $0.6861$ & $0.6862$ & $0.6844$ & $\mathbf{0.6843}$ \\\cline{2-10}
		& DR & $0.6889$ & $\mathbf{0.6886}$ & $0.6891$ & $0.6893$ & $0.6888$ & $0.6887$ & $0.6890$ & $0.6894$ \\\hline
		20 & CA & $\mathbf{0.672}$ & $0.673$ & $0.676$ & $0.675$ & $0.676$ & $0.674$ & $0.675$ & $0.675$ \\\cline{2-10}
		& DR & $\mathbf{0.681}$ & $0.684$ & $0.682$ & $0.683$ & $0.682$ & $0.682$ & $0.682$ & $0.683$ \\\hline
		50 & CA & $0.671$ & $\mathbf{0.669}$ & $0.672$ & $0.671$ & $0.674$ & $0.670$ & $0.671$ & $0.671$    \\\cline{2-10}
		& DR & $\mathbf{0.677}$ & $0.681$ & $0.679$ & $0.680$ & $0.678$ & $0.680$ & $0.683$ & $0.679$     \\\hline
		100 & CA & $0.669$ & $0.668$ & $0.669$ & $0.665$ & $0.673$ & $\mathbf{0.661}$ & $0.667$ & $0.668$ \\\cline{2-10}
		& DR & $0.6764$ & $0.6794$ & $0.6791$ & $0.6773$ & $\mathbf{0.6763}$ & $0.6798$ & $0.6787$ & $0.6782$  \\\hline
		500 & CA & $0.677$ & $0.677$ & $0.667$ & $0.678$ & $0.698$ & $\mathbf{0.661}$ & $0.670$ & $0.668$  \\\cline{2-10}
		& DR & $0.685$ & $0.689$ & $0.689$ & $\mathbf{0.681}$ & $0.696$ & $0.698$ & $0.687$ & $0.682$  \\\hline
	\end{tabular}
\end{table}

\begin{table}
	\centering
	\caption{Selected camouflage tasks}\label{tab:tasks}
	\begin{tabular}{|c|c|c|c|c|c|}
		\hline
		Secret Task & $m=2$ & $m=20$ & $m=50$ & $m=100$ & $m=500$\\\hline
		WM & OA & OA & OA & OA & OA\\\hline
		GP & 52 & 96 & 52 & 52 & 71 \\\hline
		CA & XM & BH & HB & MA & MA \\\hline
		DR & HB & BH & BH & AM & MI \\\hline
	\end{tabular}
\end{table}

\begin{table}[htb]
	\centering
	\caption{NLP solver run times}\label{tab:runtimes}
	\resizebox{\textwidth}{!}{%
		\begin{tabular}{|c|c|c|c|c|c|c|c|c|c|}
			\hline
			\multicolumn{2}{|c|}{$m=2$} & \multicolumn{2}{c|}{$m=20$} & \multicolumn{2}{c|}{$m=50$} & \multicolumn{2}{c|}{$m=100$} & \multicolumn{2}{c|}{$m=500$} \\\hline
			Dataset & Time (s) & Dataset & Time (s) & Dataset & Time (s) & Dataset & Time (s) & Dataset & Time (s) \\\hline
			WMOA & 23363 & WMOA & 29150 & WMOA & 39656 & WMOA & 32186 & WMOA & 56375 \\\hline
			GP52 & 33763 & GP96 & 65697 & GP52 & 171637 & GP52 & 161259 & GP71 & 49354 \\\hline
			CAXM & 48 & CABH & 50 & CAHB & 126 & CAMA & 86 & CAMA & 194 \\\hline
			DRHB & 44 & DRBH & 57 & DRBH & 193 & DRAM & 141 & DRMI & 205 \\\hline
		\end{tabular}
	}
\end{table}

\begin{figure}
	\centering
	\begin{subfigure}[t]{0.45\textwidth}
		\centering
		\includegraphics[width=\textwidth]{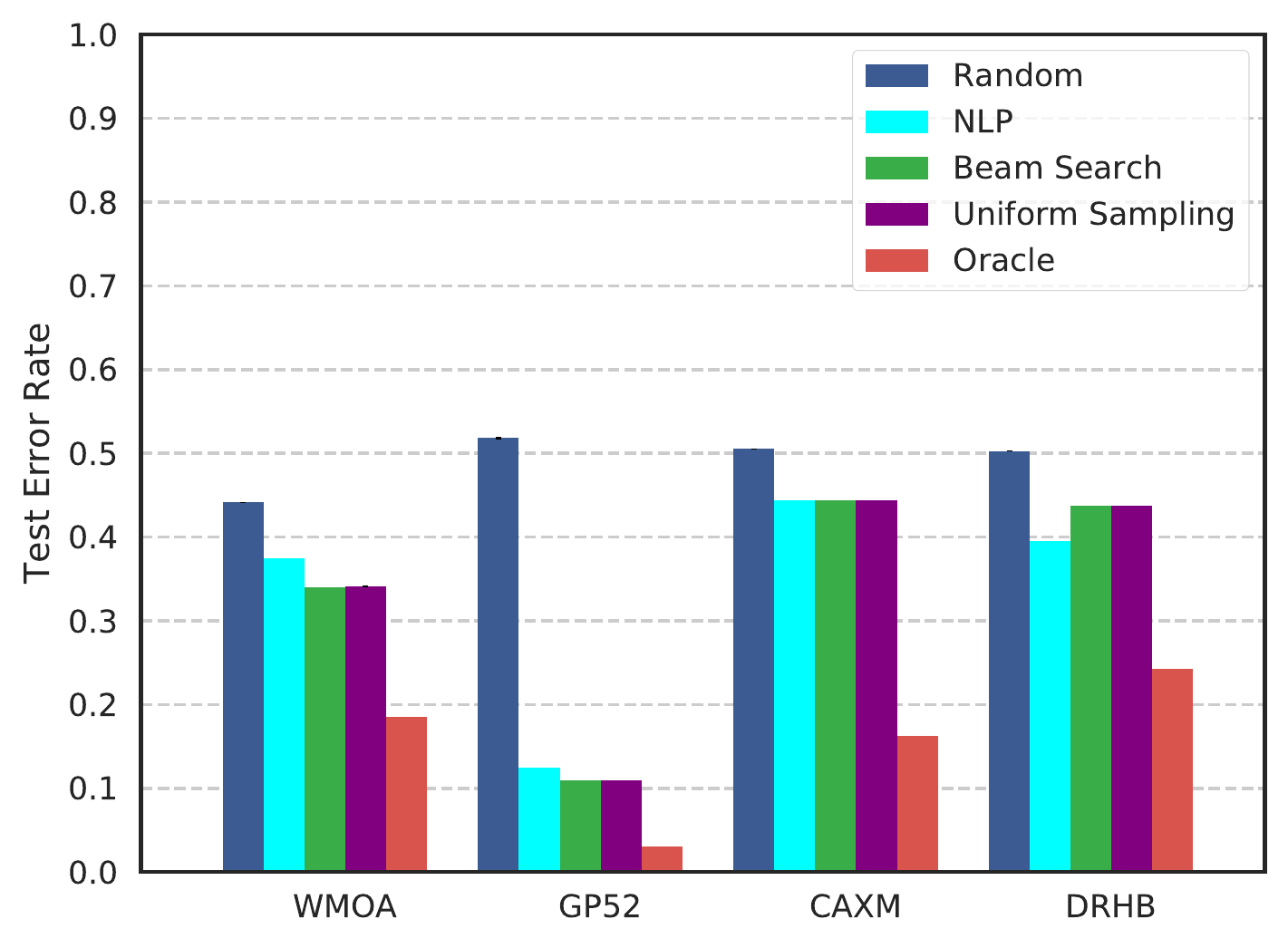}
		\caption{$m=2$}		
	\end{subfigure}
	\begin{subfigure}[t]{0.45\textwidth}
		\centering
		\includegraphics[width=\textwidth]{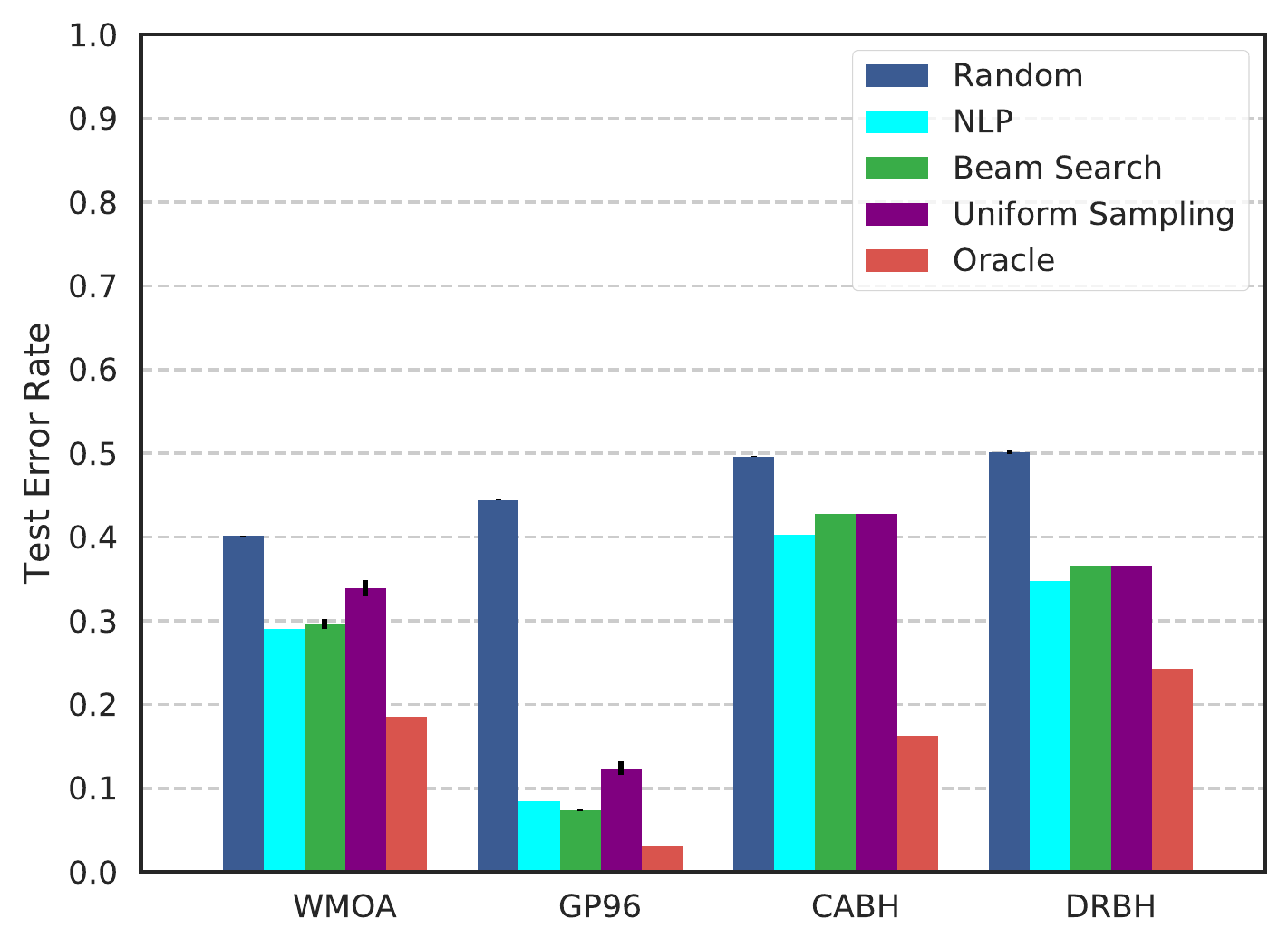}
		\caption{$m=20$}		
	\end{subfigure}
	
	\begin{subfigure}[t]{0.45\textwidth}
		\centering
		\includegraphics[width=\textwidth]{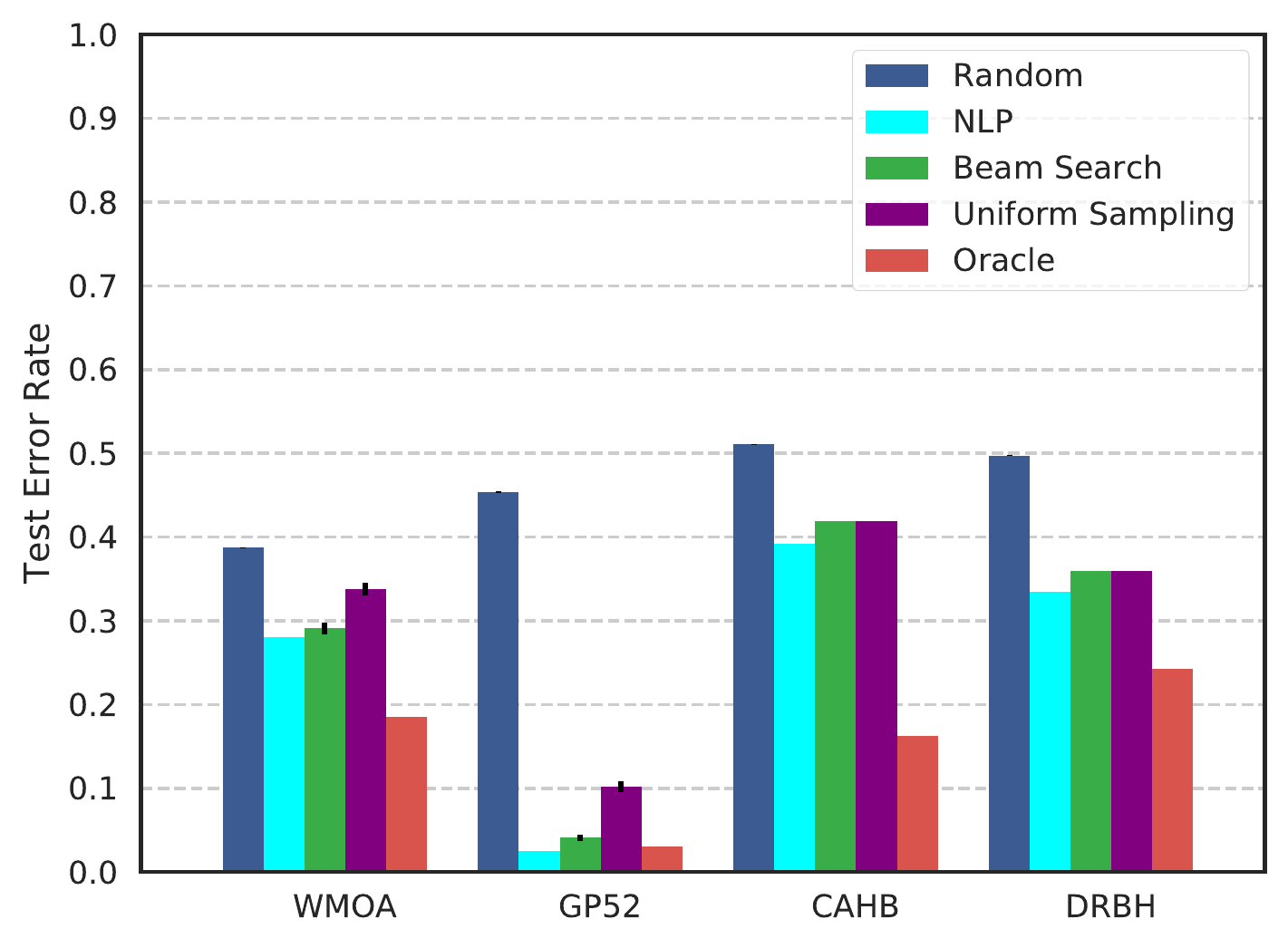}
		\caption{$m=50$}		
	\end{subfigure}
	\begin{subfigure}[t]{0.45\textwidth}
		\centering
		\includegraphics[width=\textwidth]{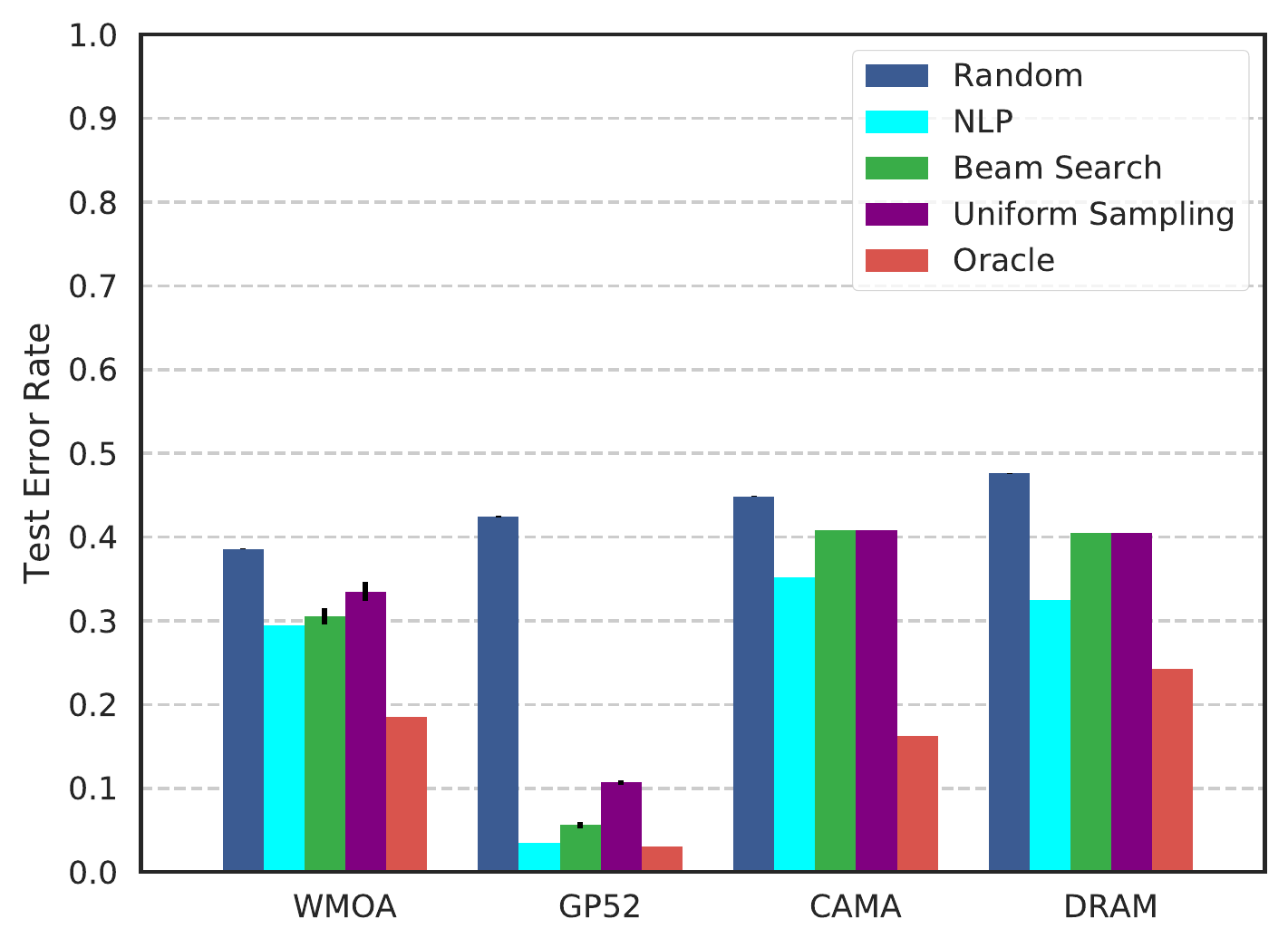}
		\caption{$m=50$}		
	\end{subfigure}
	
	\begin{subfigure}[t]{0.45\textwidth}
		\centering
		\includegraphics[width=\textwidth]{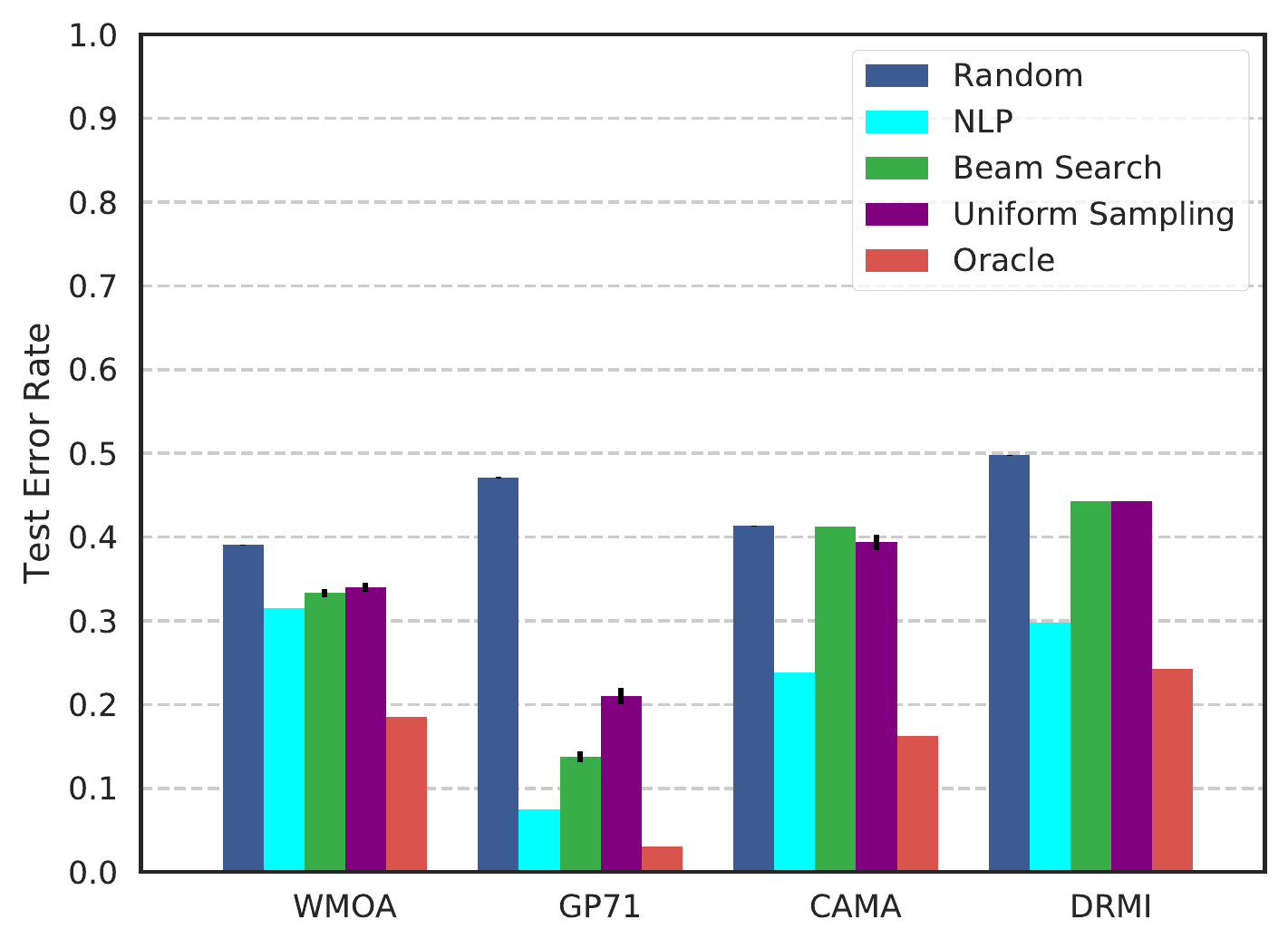}
		\caption{$m=50$}		
	\end{subfigure}
	\caption{Test error rates found by solving the camouflage framework. We also show random and oracle error for comparison. Error bars are also shown. All three solvers were run for the same amount of time.}~\label{fig:lr_results_same}
\end{figure}

\begin{figure}
	\centering
	\begin{subfigure}[t]{0.45\textwidth}
		\centering
		\includegraphics[width=\textwidth]{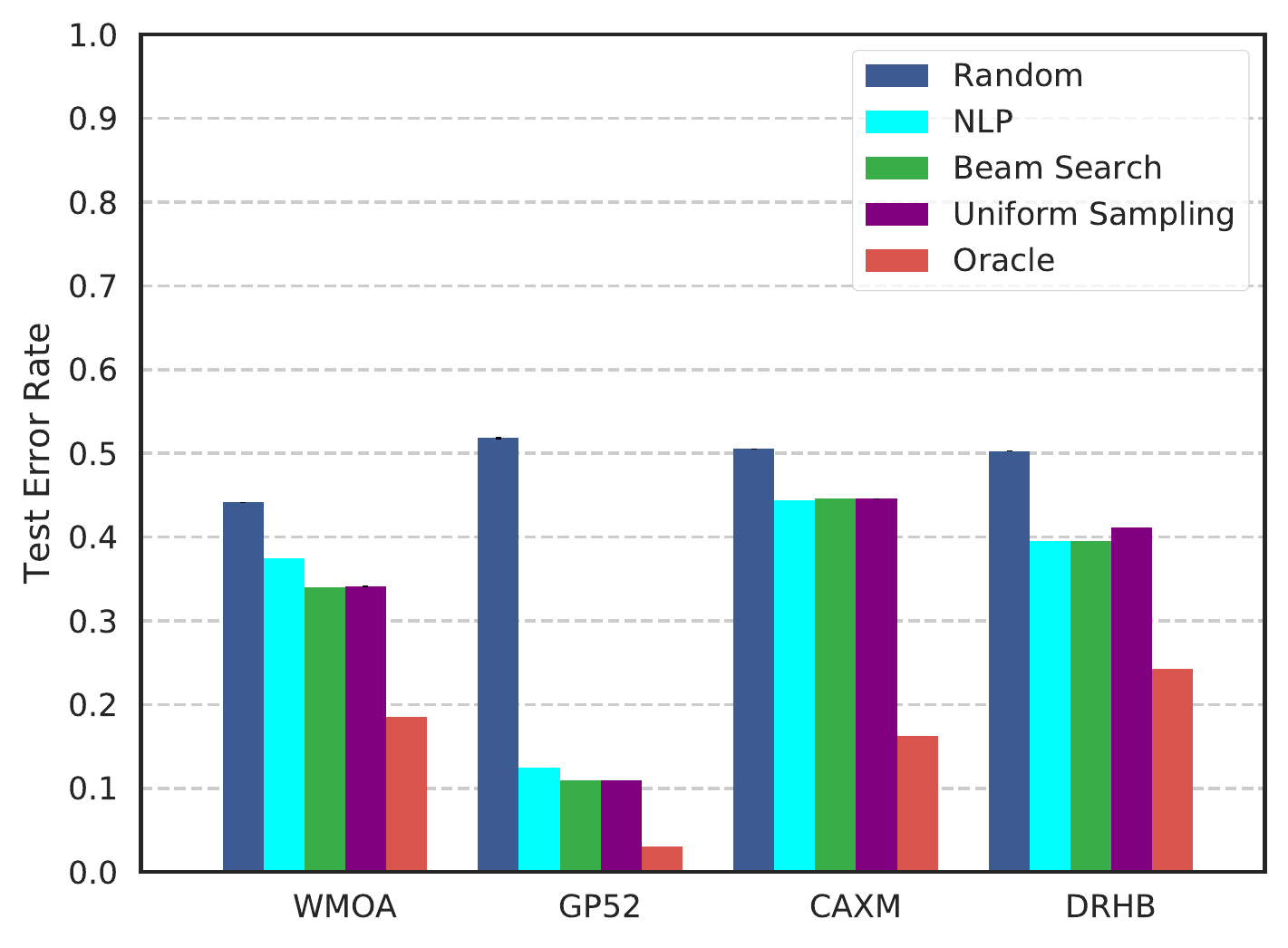}
		\caption{$m=2$}		
	\end{subfigure}
	\begin{subfigure}[t]{0.45\textwidth}
		\centering
		\includegraphics[width=\textwidth]{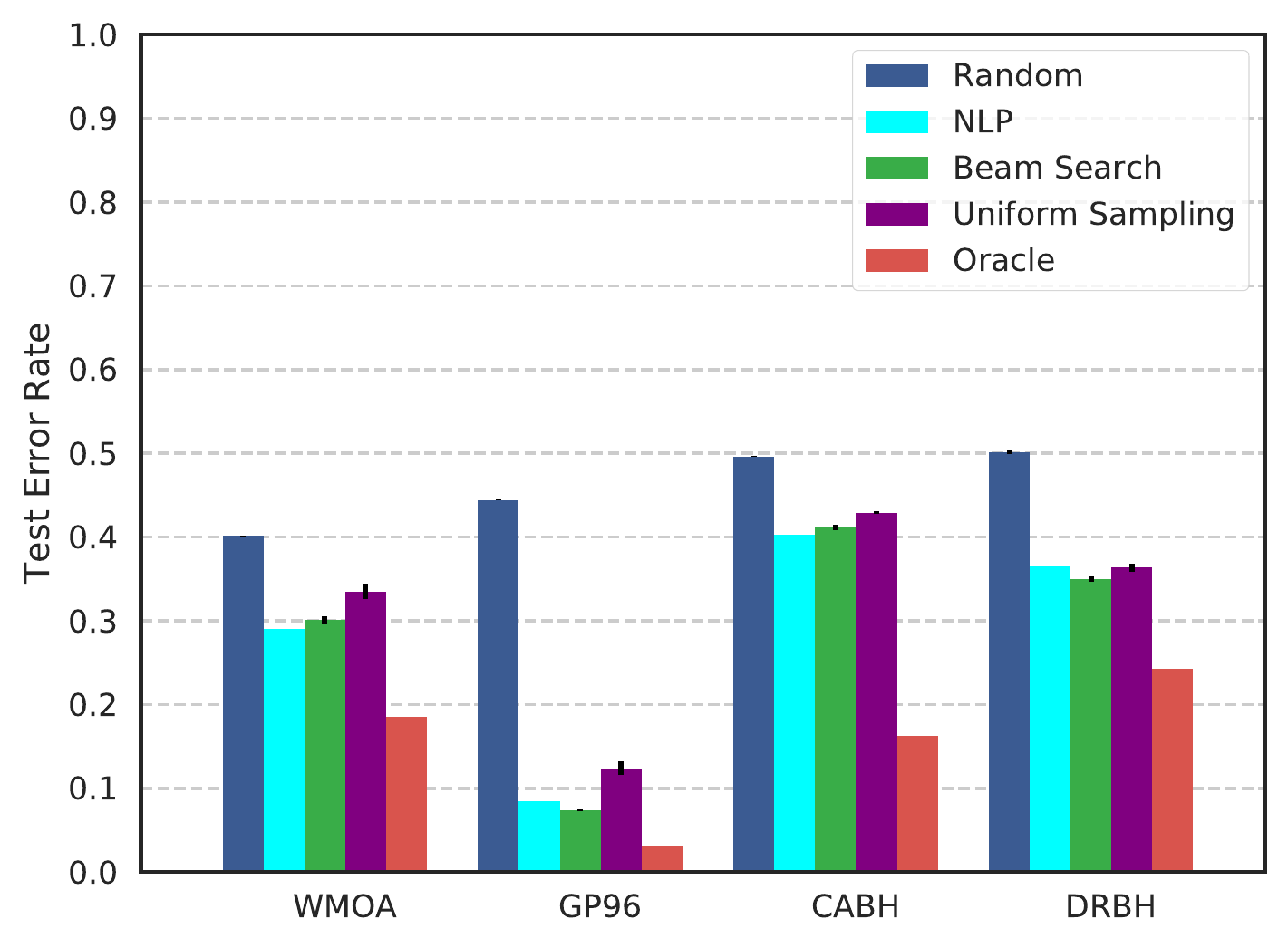}
		\caption{$m=20$}		
	\end{subfigure}
	
	\begin{subfigure}[t]{0.45\textwidth}
		\centering
		\includegraphics[width=\textwidth]{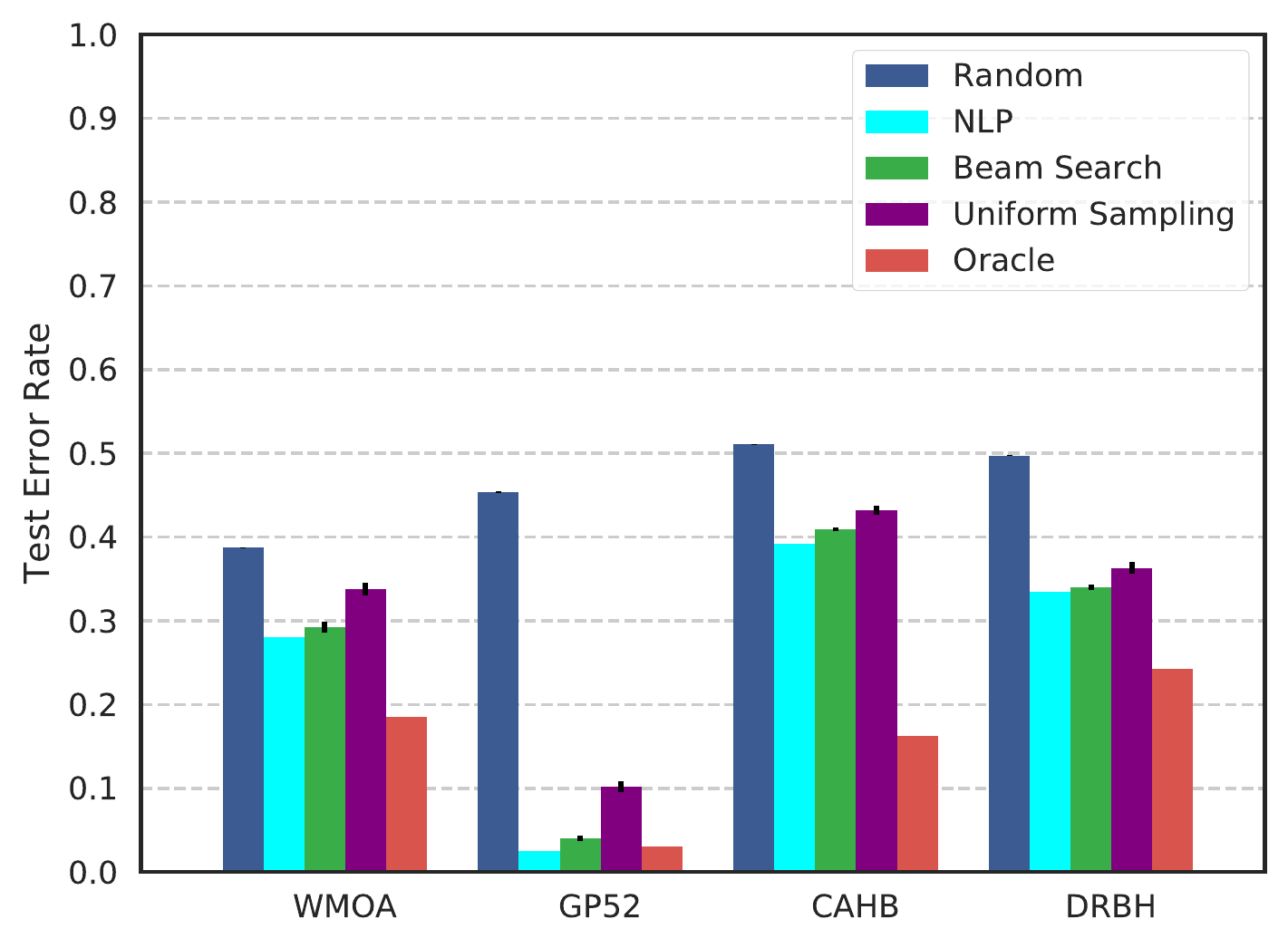}
		\caption{$m=50$}		
	\end{subfigure}
	\begin{subfigure}[t]{0.45\textwidth}
		\centering
		\includegraphics[width=\textwidth]{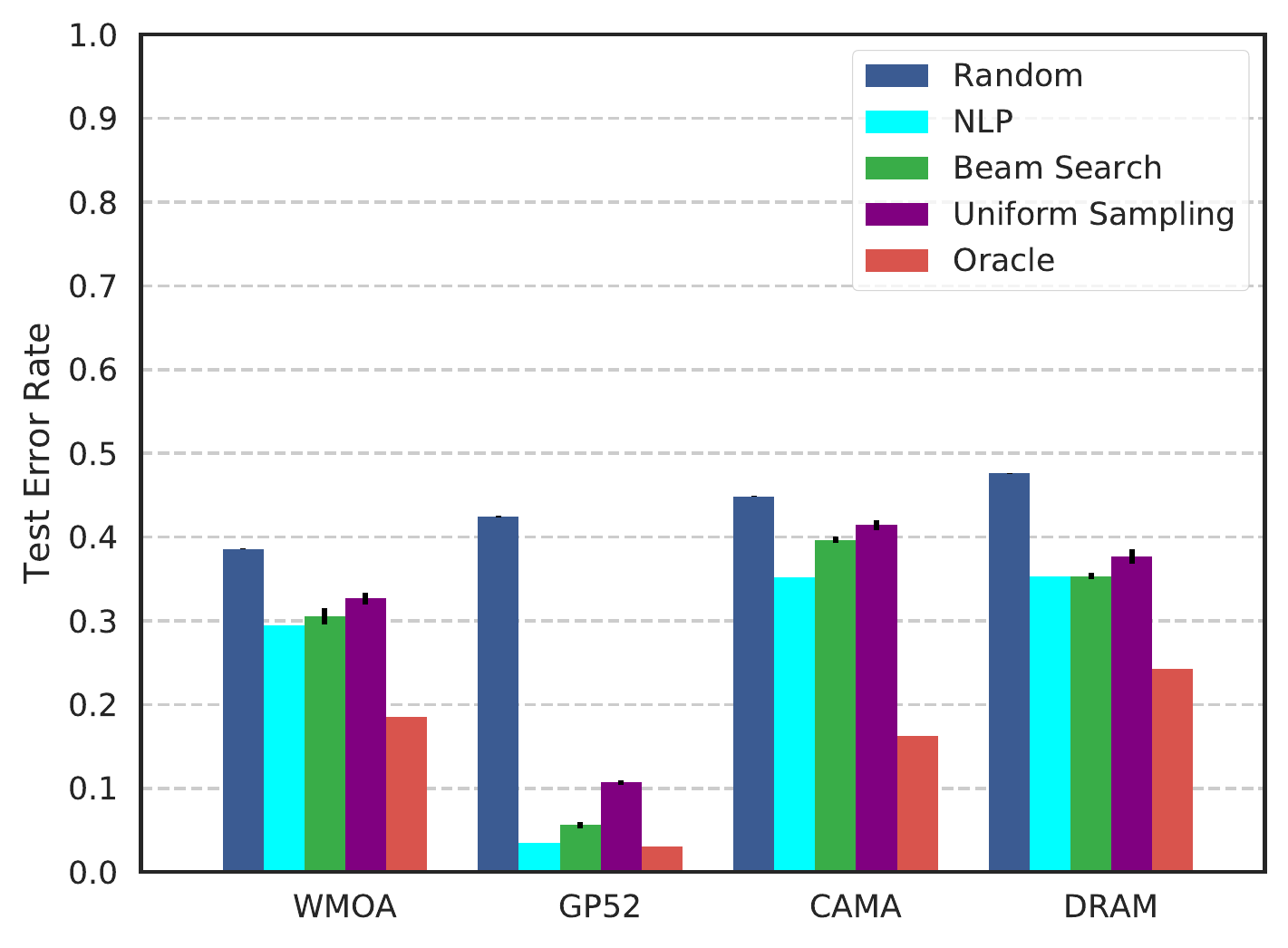}
		\caption{$m=50$}		
	\end{subfigure}
	
	\begin{subfigure}[t]{0.45\textwidth}
		\centering
		\includegraphics[width=\textwidth]{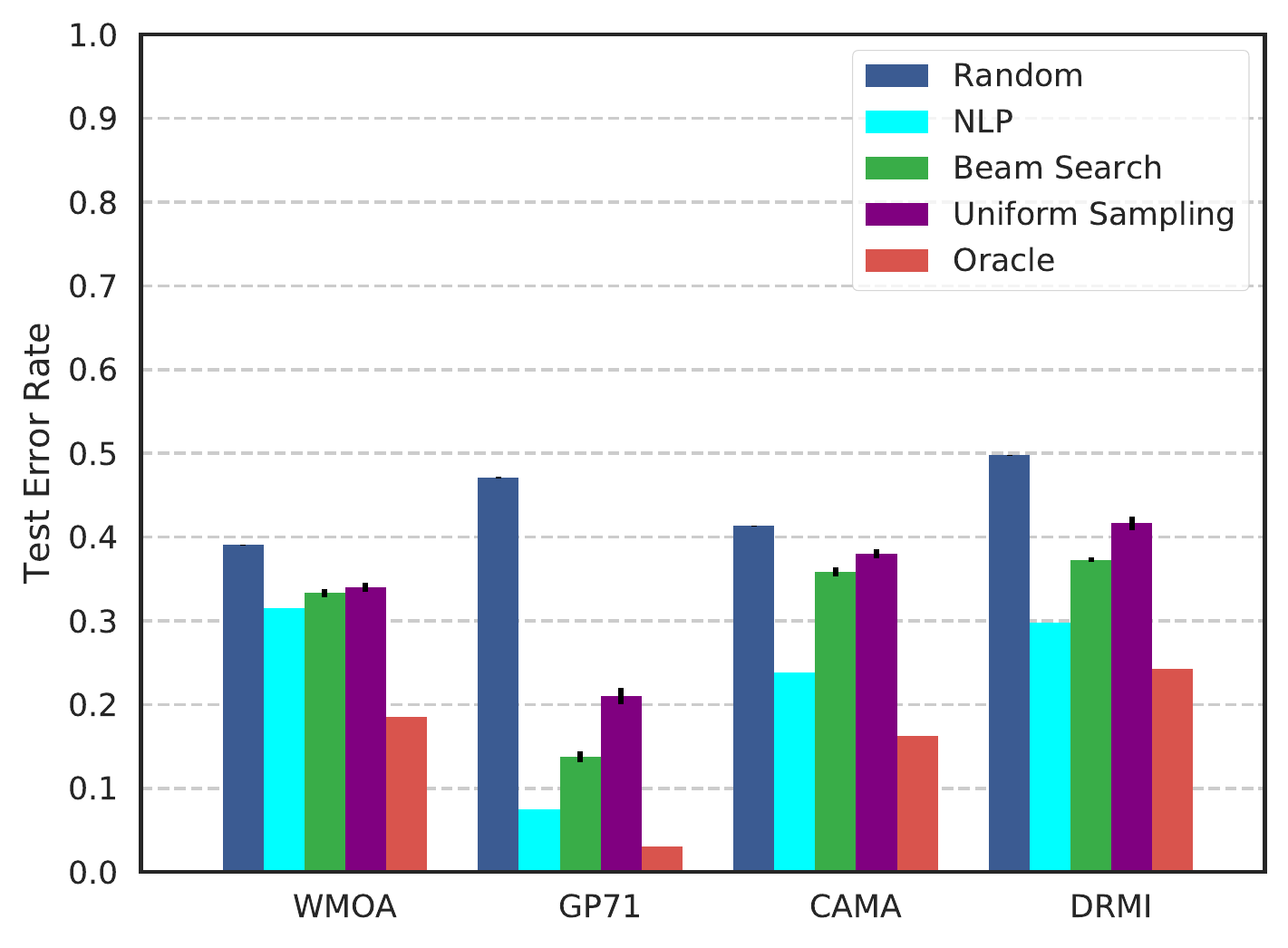}
		\caption{$m=50$}		
	\end{subfigure}
	\caption{Test error rates found by solving the camouflage framework. We also show random and oracle error for comparison. Error bars are also shown. The Beam Search and Uniform Search solvers were run for an additional two hours more than the NLP solver run time.}~\label{fig:lr_results_extended}
\end{figure}

\subsection{Results}
The run times for the NLP solver are presented in Table~\ref{tab:runtimes}.
It can be seen from the table that due to the high dimensionality, the NLP solver takes a much longer time to find a solution for the image tasks.
We present our results for all three solvers in Figure~\ref{fig:lr_results_same} and Figure~\ref{fig:lr_results_extended}. 
Figure~\ref{fig:lr_results_same} shows the results when all three Alices run their solvers for the same amount of time.
In Figure~\ref{fig:lr_results_extended} we show the results when Beam Search and Uniform Search solvers are run for an additional two hours more than the NLP solver run time.
For the text secret tasks, Alice could not find a better camouflaged training set using either beam search or uniform sampling than the one found during the initial run of uniform sampling (with a total budget of 80,000).
To explore the sensitivity of beam search and uniform sampling regarding the time budget, we ran both solvers for an additional two hours.
But the results only improved marginally (Figure~\ref{fig:lr_results_extended}).
We observe that Alice, using any of the three solvers can find much better camouflage training sets than random and in many cases approach oracle performance.
Note that Alice's solutions do not trigger Eve's suspicion function.
This shows that such subterfuges are plausible in practice and can actually yield good results from Alice's point of view.
We note that Alice yields the best results when $m=50$ and $m=500$ respectively for image and text tasks in most of the experiments. 
But this may not hold for larger values of $m$ e.g., when $m$ is equal to the size of the camouflage pool.
We plan to run further experiments to understand the effect of $m$.

Figure~\ref{fig:example} shows the result of WMOA when Bob's learner is logistic regression and the solver is NLP $(m=20)$.
Visually, the camouflaged training set $D$ bears no resemblance to the secret training set $D_S$.
This is true for the text camouflage experiments as well, where articles in the camouflaged training sets have no obvious semantic connection to the secret task. 
See Table~\ref{tab:CABH} for results on the text experiment CABH.
This is indeed bad news for human Eves: not only did camouflage fooled MMD detector, it will also likely fool human inspectors.

\begin{table}[htb]
	\centering
	
	\caption{Camouflage results for the CABH experiment with $m=20$ for the NLP solver}
	\label{tab:CABH}
	
	\resizebox{\textwidth}{!}{%
		\begin{tabular}{|l|l|l|l|}
			\hline
			\multicolumn{2}{|c|}{\textbf{Sample of Secret Set}} & \multicolumn{2}{|c|}{\textbf{Sample of Camouflaged Training Set}} \\\hline
			\multicolumn{1}{|c|}{\textbf{Class}} & \multicolumn{1}{|c|}{\textbf{Article}} & \multicolumn{1}{|c|}{\textbf{Class}} & \multicolumn{1}{|c|}{\textbf{Article}} \\\hline
			Christianity & $\ldots$Christ that often causes christians to be very & Baseball & $\ldots$Boys, hats off to any Cubs fan who can actually\\ 
			&  critical of themselves and other christinas. We$\ldots$ & &  muster up the courage to put down Braves fans. I$\ldots$\\\cline{2-2}\cline{4-4}
			& $\ldots$I've heard it said that the accounts we have   & & $\ldots$ NPR's Morning Edition aired a report this morning\\
			& of Christs life and ministry in the Gospels were$\ldots$ & &  to get (4/19) on Hispanic/Latin American players in MLB$\ldots$ \\\hline
			Atheism& $\ldots$This article attempts to provide a general  & Hockey & $\ldots$ Would Kevin Dineen play for the Miami Colons???   \\
			& introduction to atheism. Whilst I
			have tried to be$\ldots$ & & As a Flyers fan, I resent you making Kevin Dineen$\ldots$\\\cline{2-2}\cline{4-4}
			& $\ldots$Science is wonderful at answering most of our  & & $\ldots$Good point - there haven't even been any recent posts 
			\\
			& questions. I'm not the type
			to question scientific$\ldots$ & & about ULf! Secretly, I'm convinced that he is responsible $\ldots$\\\hline
		\end{tabular}}
\end{table}

%% file: relwork.tex
Concealing the existence of messages is known as steganography. 
One illustration of steganography (first presented in 1983 in \cite{simmons1984prisoners}) is where prisoners Alice and Bob wish to devise an escape plan. 
All their communication is observed by the adversary (the warden, Eve) who will thwart their plan as soon as she detects any sign of hidden message. 

Steganography has multiple real-world applications including secret communication~\cite{zhang2009image}, feature tagging elements~\cite{maganbhaistudy}, and copyright protection~\cite{maganbhaistudy}.
Although many different data formats can be used for steganography, images~\cite{queirolo2011steganography,johnson1998exploring} are by far the most popular format due to their popularity on the internet and the fact that they are rich with noise-insensitive information.
Image steganography can be broadly classified into spatial domain, transform domain, spread spectrum and model based \cite{singh2014survey}, and has been thoroughly studied.
On the other side, steganalysis is the study of detecting the existence of hidden messages (using steganography).
Identifying such messages in text by looking at patterns in texts, odd language and unusual white space was explored in~\cite{chandramouli2002mathematical}.
The authors of 
\cite{fridrich2004feature,ker2005steganalysis,queirolo2011steganography} 
 explore the detection of hidden messages in images.

A study of steganography from a complexity-theoretic point of view is presented in~\cite{hopper2002provably,reyzin2003more}.
An information-theoretic model for such a setup is presented in~\cite{cachin1998information}. 
This complexity-theoretic security notion is similar to modern cryptography and they try to define a secure stegosystem such that the stegotext is computationally indistinguishable from the covertext.
In such a scenario a new term called steganographic secrecy of stegosystem is introduced which is defined as the inability of a polynomial-time adversary (Eve) to distinguish between observed distributions of unaltered covertext and stegotexts.
To the best of our knowledge, steganographic techniques have not been used in the domain of training sets for machine learning models.

Steganography is often confused with cryptography~\cite{katz1996handbook,van2014encyclopedia}, however the goal of these two systems are completely different. 
The goal of cryptography is to ensure confidentiality of data in communication and storage processes.
Hiding the existence of sensitive data is not the end goal here (unlike steganography). 
According to Kerckhoffs's principle~\cite{kerckhoffs18831,kerckhoffs18832}, this confidentiality must not rely on the obfuscation of the encoding scheme, but only on the secrecy of the decryption key.


One particular branch of cryptography we highlight is {homomorphic encryption}~\cite{rivest1978data}.
Consider a situation where you seek to delegate some computation to another computer (e.g., using a cloud computation service to a perform machine learning task).
You would like to utilize their computation power, but you do not trust them with your private data.
Homomorphic encryption allows a method by which you can encrypt your data prior to sending it.
The untrusted computer will then perform its operations on the encrypted data, returning to you the result (e.g., a learned model).
You then decrypt the result, yielding what the remote computer would have computed had you provided your original (unencrypted) data.
A homomorphic cryptosystem which supports arbitrary computation on ciphertexts is known as fully homomorphic encryption (FHE).
The first plausible construction of such a system was proposed in~\cite{smart2010fully}. 
This scheme supports both addition and multiplication operations on ciphertexts, which in turn makes possible to construct circuits for arbitrary computations.
Some second generation solutions were proposed in~\cite{brakerski2014leveled,brakerski2012fully,lopez2012fly,gentry2013homomorphic}.

In our setting, encryption (homomorphic or otherwise) is not enough to solve Alice's task.
After Alice has transmitted her data to Bob, Bob learns a model.
Alice's goal is not only for Eve to not know the model (which could easily be achieved by Alice simply sending an encrypted model), but also for Eve not to be suspicious.
Eve believes that Alice is drawing data points i.i.d. from some distribution and thus data encrypted by standard methods will cause alarm.
We do note that the relatively new method of ``honey encryption''~\cite{juels2014honey} may be a useful alternative approach for Alice, which we leave as future work.

The idea of constructing a dataset keeping a particular machine learning algorithm and a target model in mind is known as machine teaching. 
Machine teaching is the inverse of machine learning and has applications in various fields~\cite{zhang2018training,liu2016teaching}. 
In particular, machine teaching has applications in the domain of adversarial learning which studies the use of machine learning in security-sensitive domains.
Numerous attacks against various machine learners have been explored, highlighting the security ramifications of using machine learning in practice \cite{huang2011adversarial,barreno2006can,barreno2010security,dalvi2004adversarial,laskov2009framework,tan2002undermining}.

In the work presented herein, Alice can be thought of as ``attacking'' the learner Bob, in that she aims to provide a dataset which causes Bob to learn a model with parituclar properties.
We highlight how this differs from the classical adversarial learning framework in two ways.
First, Alice is not perturbing an existing training set, but rather generating one.
Thus, this is more akin to the Machine Teaching framework.
Second is the presence of Eve.
Namely, Alice is trying not only to affect Bob's resulting model, but also to hide her involvement from a third party eavesdropper.
In spirit, this is similar to the adversarial learning work performed on intrusion detection systems~\cite{kloft2007poisoning,kloft2011online}.
In terms of the details of the mathematics, our framework and strategies for solving Alice's optimization problem more closely follow~\cite{mei2015using}.

Within adversarial machine learning, a line of research has posed the problem of learning in the presence of adversaries in game theoretic contexts~\cite{liu2009game,bruckner2012static,dalvi2004adversarial,bruckner2009nash,bruckner2011stackelberg,hardt2016strategic,bruckner2011stackelberg}. 
\cite{dalvi2004adversarial,letchford2013optimal,alfeld2017explicit} specifically address a learner's defense strategy in various contexts.
Randomization has also been explored as a method of defense~\cite{bulo2016randomized,vorobeychik2014optimal}, as well as in the context of machine teaching \cite{balbach2006teaching}.
Our work contributes to this conversation as Eve can be seen as a form of defense for Bob.

%% file: discussions.tex
We introduced the training set camouflage setting where a carefully constructed training set can be sent over an open channel with the intention of training a machine learner on a secret classification task.
Using this framework, an agent can hide the intention of the secret task from a third party observer.
Our experimental results show that training set camouflage is indeed a plausible threat.
We present three approaches to solve the optimization problem.
We observe that all three solvers perform well but both NLP and beam search outperform uniform sampling in all cases.
The NLP solver often performs a bit better than beam search.
This suggests that for the logistic regression learner NLP is Alice's preferred solver of choice.
However, the NLP solver cannot be applied to all possible learners (non-convexity prevents the application of KKT conditions).
Thus in such cases beam search becomes the preferred solver.

We note that $\mmd$ is stronger with larger sample sizes.
It will be harder for Alice to fool Eve given a large camouflage pool $C$ and also if she is forced to select a large camouflaged training set $D$. 
$\mmd$ is also stronger with smaller feature dimensions~\cite{gretton2012kernel}.
Also, it is harder for Alice to fool Eve if she increases the value of $\alpha$.
Since $\alpha$ is the upper bound of the probability of the Type I error for the null hypothesis i.e., the camouflage pool and camouflaged training set come from the same distribution, increasing $\alpha$ allows Eve to become more suspicious.
As future work we plan to devise defensive strategies against Alice.
In such scenarios it is advisable to assume that Eve's detection function is known to the attacker (Kerckhoffs's principle~\cite{kerckhoffs18831,kerckhoffs18832}) which we make here.

We note that camouflage seems easier for Alice to do if the cover task is in fact somewhat confusable, presumably because she can generate different decision boundaries by picking from overlapping camouflage items.
This can be imagined easily in the 2D case with two overlapping point clouds forming the cover task.
In such a scenario any separable secret task (no overlap between the secret task instances) can be taught to Bob by Alice. 
One interesting open question is whether there is a \textit{universal} cover task for all secret tasks. 
We also note that achieving Alice's goal becomes much harder in the multi-class setting as finding a cover task becomes more challenging.
	
As mentioned previously, Bob fixed his learning hyperparameters (e.g., regularization parameter of the logistic regression).
This was done for speed.
However, nothing prevents Bob from using cross validation~\cite{kohavi1995study}. 
Cross validation is popular technique used in machine learning where the learner is trained multiple times on different subsets of the whole training set to tune the hyperparameters of the learner.
Alice would simply emulate the same cross validation while optimizing the camouflaged training set.
This can be easily done in beam search and uniform sampling, at the cost of more computation.
Unfortunately significant modifications will be required for NLP.
	
Also, the loss function $\ell$ used by Alice and Bob is the same, as seen in the upper and lower optimization problems in~\eqref{eq:MMDConstraint}.
It is straightforward to allow different losses.  For example, Bob may learn with the logistic loss since it is a standard learner, while Alice uses 0-1 loss to directly optimize Bob's accuracy.
	
	
We note that training set camouflage can be extended to cross modality correspondence, e.g., use an image camouflage pool while the secret classification task is to classify text articles. 
Alice and Bob can communicate via the private channel to establish the correspondance between images features and text features. 
Another possible way to extend the camouflage pool is to allow perturbed instances as well.

%% file: mmd.tex
One critical component of our camouflage framework is Eve's detection function \(\suspicion\) ---  how she determines if a training set is suspicious or not.
Eve's detection function is a two-sample test as its goal is to discern if the two sets $\cand, D$ are drawn from the same distribution or not.
In what follows we discuss using Maximum Mean Discrepancy ({\mmd})~\cite{gretton2012kernel} as Eve's detection function, as we do in our experiments.
{\mmd} is a widely used two-sample test~\cite{dziugaite2015training}, but, of course other detection functions can be used in~(\(\ref{aliceopti}\)).

We first review basic $\mmd$ following~\cite{gretton2012kernel}.
Let $p$ and $p'$ be two Borel probability measures defined on a topological space $\mathcal{Z}$.
Given a class of functions $\fspace$ such that $f:\mathcal{Z}\mapsto\R, f\in\fspace$, $\mmd$ is defined as 
\begin{equation}
	\mmd(p,p')=\sup_{f\in\fspace}(E_\bfz[f(\bfz)]-E_{\bfz'}[f(\bfz')])
\end{equation}
Any unit ball in a reproducing kernel Hilbert space (RKHS) can be used as the function class $\fspace$ if the kernel is universal (e.g., Gaussian and Laplace kernels~\cite{steinwart2001influence}).
Using this function space, $\mmd$ is a metric.
This means $\mmd(p,p') = 0 \Leftrightarrow p = p'$.

Computing $\mmd$ requires the expectations to be known, which generally, is not the case in practice.
We obtain an empirical estimation by replacing the population expectations with empirical mean computed on 
i.i.d. samples $Z=\{\bfz_1,\ldots,\bfz_\candSize\}$ and $Z'=\{\bfz'_1,\ldots,\bfz'_\trainSize\}$ from $p$ and $p'$, respectively. 
We define
\begin{equation}
\mmd(Z,Z') =  \left[\frac{1}{\candSize^2}\sum_{i,j=1}^\candSize k(\bfz_i,\bfz_j)-\nonumber\frac{2}{\candSize\trainSize}\sum_{i,j=1}^{\candSize,\trainSize}k(\bfz_i,\bfz'_j) +\frac{1}{\trainSize^2}\sum_{i,j=1}^\trainSize k(\bfz'_i,\bfz'_j)\right]^\frac{1}{2} 
\end{equation}
where $k$ is the kernel of the RKHS.
Let $d=\vert \mmd(Z,Z')-\mmd(p,p')\vert$.
Gretton {\em et. al.} show that 
\begin{equation}
P\left(d > 2 \left(\sqrt{\frac{K}{n}} + \sqrt{\frac{K}{m}}\right) + \epsilon \right) \le 2 e^{-\frac{\epsilon^2nm}{2K(n+m)}}
\end{equation}
where $K$ is an upperbound on the kernel values.

We convert the above bound into a one-sided hypothesis testing procedure.
Under the null hypothesis $p=p'$ we have $\mmd(p,p')=0$. 
We consider positive deviations of $\mmd(Z,Z')$ from $\mmd(p,p')$.
Equating the RHS with $\alpha$ (probability of incorrectly stating $p\ne p'$ also known as the type I error) gives a hypothesis test of level-$\alpha$, where solving $\epsilon$ as a function of $\alpha$ gives
$
\alpha = e^{-\frac{\epsilon^2nm}{2K(n+m)}} \Rightarrow \epsilon = \sqrt{\frac{2K(n+m)}{nm}\log\frac{1}{\alpha}}
$.
We retain the null hypothesis if
\begin{equation}
\mmd(Z,Z') - T <0
\end{equation}
where the threshold is
\begin{equation}
T = 2 \left(\sqrt{\frac{K}{n}} + \sqrt{\frac{K}{m}}\right) + \sqrt{\frac{2K(n+m)}{nm}\log\frac{1}{\alpha}}
\end{equation}
This also defines Eve's detection function ($\suspicion(\cand,\trset)$) at level-$\alpha$: 
\begin{equation}
\suspicion(\cand,\trset)\equiv\mmd(\cand,\trset) - T.
\end{equation}
If $\suspicion(\cand,\trset) \ge 0$ then Eve realizes that $\trset$ is not drawn i.i.d. from $\mathbb{\cdist}_{(\bfx, y)}$ and flags it as suspicious.




For all our experiments Eve used the RBF kernel $k(\bfz_i, \bfz_j) = \exp\left(-\frac{\Vert \bfz_i - \bfz_j \Vert^2}{2\sigma^2}\right)$. 
Eve set $\sigma$ to be the median distance between points in the camouflage pool as proposed in ~\cite{gretton2012kernel}. 
Eve also included the scaled class label as a feature dimension: $[\bfx_i, c \mathbbm{1}\{y_i=1\}]$ where
$c=\max_{k,l\text{ such that } y_k = y_l} \Vert \bfx_{k} - \bfx_{l}\Vert$
and $\mathbbm{1}\{\cdot\}$ is the indicator function.
This augmented feature enables Eve to monitor both features and labels. 
When using the NLP solver Alice only has to consider instances from camouflage pool. 
She calculated $\mmd$ in the following manner:
\begin{align}
\mmd_b(Z, b_1,\ldots,b_{|Z|})=&\left[\frac{1}{\candSize^2}\sum_{i,j=1}^\candSize k(\bfz_i,\bfz_j) - \frac{2}{n\sum_{i=1}^{\candSize} b_i}\sum_{i,j=1}^{\candSize} b_ik(\bfz_i,\bfz_j)\right.\nonumber\\
& \left.+ \frac{1}{(\sum_{i=1}^{n}b_i)^2} \sum_{i,j=1}^{n} b_ib_jk(\bfz_i,\bfz_j)\right]^\frac{1}{2}
\end{align}